\documentclass[preprints,article,submit,pdftex,moreauthors]{Definitions/mdpi} 
\firstpage{1} 
\pubvolume{1}
\issuenum{1}
\articlenumber{0}
\pubyear{2025}
\copyrightyear{2025}
\datereceived{ } 
\daterevised{ } 
\dateaccepted{ } 
\datepublished{ } 
\hreflink{https://doi.org/} 

\usepackage{amsmath}

\Title{Modelling variability of the immunity build-up and waning following RNA-based vaccination}

\TitleCitation{Modelling variability of the immunity build-up and waning following RNA-based vaccination}

\Author{Juan Magalang $^{1,2}$*\orcidA{}, Tyll Krueger $^{3}$ and Joerg Galle $^{4}$*}

\AuthorNames{Juan Magalang, Tyll Krueger and Joerg Galle}

\isAPAStyle{%
       \AuthorCitation{Magalang, J., Krueger, T., \& Galle, J.}
         }{%
        \isChicagoStyle{%
        \AuthorCitation{Magalang, Juan, Tyll Krueger, and Joerg Galle.}
        }{
        \AuthorCitation{Magalang, J.; Krueger, T.; Galle, J.}
        }
}

\address{%
$^{1}$ \quad Department of Visceral Surgery and Medicine, Inselspital, Bern University Hospital, University of Bern, Murtenstrasse 35, Bern 3008, Switzerland\\
$^{2}$ \quad Institute of Mathematical Statistics and Actuarial Science, University of Bern, Alpeneggstrasse 22, Bern 3012, Switzerland\\
$^{3}$ \quad Faculty of Information and Communication Technology, Wroclaw University of Science and Technology, Janiszewskiego 11-17, 50-372 Wrocław, Poland\\
$^{4}$ \quad Interdisciplinary Centre for Bioinformatics (IZBI), Leipzig University, Haertelstr. 16-18, 04107 Leipzig, Germany}

\corres{Correspondence: juan.magalang@unibe.ch, galle@izbi.uni-leipzig.de}

\abstract{RNA-based vaccination has been broadly applied in the COVID pandemic. A characteristic of the immunization was fast waning immunity. However, the time scale of this process varied considerable for virus subtypes and among individuals. Understanding the origin of this variability is crucial in order to improve future vaccination strategies. Here, we introduce a mathematical model of RNA-based vaccination and the kinetics of the induced immune response. In the model, antigens produced following vaccination rise an immune response leading to germinal center reactions and accordingly B-cell differentiation into memory B-cells and plasma cells. In a negative feedback loop, the antibodies synthesized by newly specified plasma cells shut down the germinal center reaction as well as antigen-induced differentiation of memory B-cell into plasma cells. This limits the build-up of long-lasting immunity and thus is accompanied by fast waning immunity. The detailed data available on infection with and vaccination against SARS-CoV-2 enabled computational simulation of essential processes of the immune response. By simulation, we analyzed to which extent a single or double dose vaccination provides protection against infection. We find that variability of the immune response in individuals, originating e.g. in different immune cell densities, results in a broad log-normal–like distribution of the vaccine-induced protection times that peaks around 100 days. Protection times decrease for virus variants with mutated antibody binding sites or increased replication rates. Independent of these virus specifics, our simulations suggest optimal timing of a second dose about 5 weeks after the first in agreement with clinical trials.
}

\keyword{Mathematical modelling, RNA vaccines, immune response} 

\begin{document}


\section{Introduction}

The immune response enables our body to defend itself against pathogens including viruses and bacteria. Mathematical models describing the kinetics of the immune response against infections with them are well established. In many cases, they combine aspects of the innate and adaptive response (e.g. in case of influenza, \cite{Lee2009}). While the innate response is directed against the spreading infection, the adaptive response mainly serves to protect the body from potential future infections. The effectivity of the latter comes into focus when vaccination is applied. A core component of the adaptive response are germinal center reactions enabling the specification of cells capable of producing antibodies that can efficiently neutralize the pathogens. This process can provide lifelong protection against future infections. Modeling these reactions has attracted increasing attention in the last years \cite{Molari2020, MeyerHermann2021}. 

The source of antibodies are plasma cells \cite{Akkaya2019}. Two types of these cells are typically specified during immune responses against a pathogen V, short- and long-living plasma cells. Short-living plasma cells vanish over several weeks. They become newly induced during repeated infection with V. Their long-living counterparts are resident in the bone marrow and permanently secret V-specific antibodies. Thus, they are most efficient in suppressing repeated infections with V.

The emergence of SARS-CoV-2 led to the first-time broad application of RNA-based vaccination.  Several mathematical models have been introduced that provide insight into the principles of the immune response on both infection by the virus and vaccination against it (e.g. \cite{Leon2023}).  In summer 2021, it became obvious that the immunization reached by vaccination significantly decreases on the time scale of several months. Starting with reports about breakthrough infections from Israel \cite{Kustin2021}, an increasing number of studies provided insights into this waning process \cite{Menegale2023}. Meanwhile, related effects have been integrated into mathematical models aiming at forecasts of infection numbers \cite{Bicher2022}. Some of them link the level of remaining protection and control measures such as antibody concentrations in the serum \cite{Uwamino2023, Russell2025}. The cellular and molecular factors that control individual waning of protection against infection remain largely open. A potential explanation of the fast waning is missing specification of long-living plasma cells following vaccination \cite{Nguyen2024}. 

Here, we introduce a mathematical model of RNA-based vaccination against a virus V and the kinetics of the induced immune response. We focus on the adaptive immune response and the processes leading to production of antibodies against V-specific antigens. We quantify the immunization reached by the vaccination by the concentration of the antibodies and their capability to bind the antigen. We observe waning immunity and identify potential sources of this process. 

\section{Basic assumptions}

\begin{figure}[!htb]
    \centering
    \includegraphics[width=\linewidth]{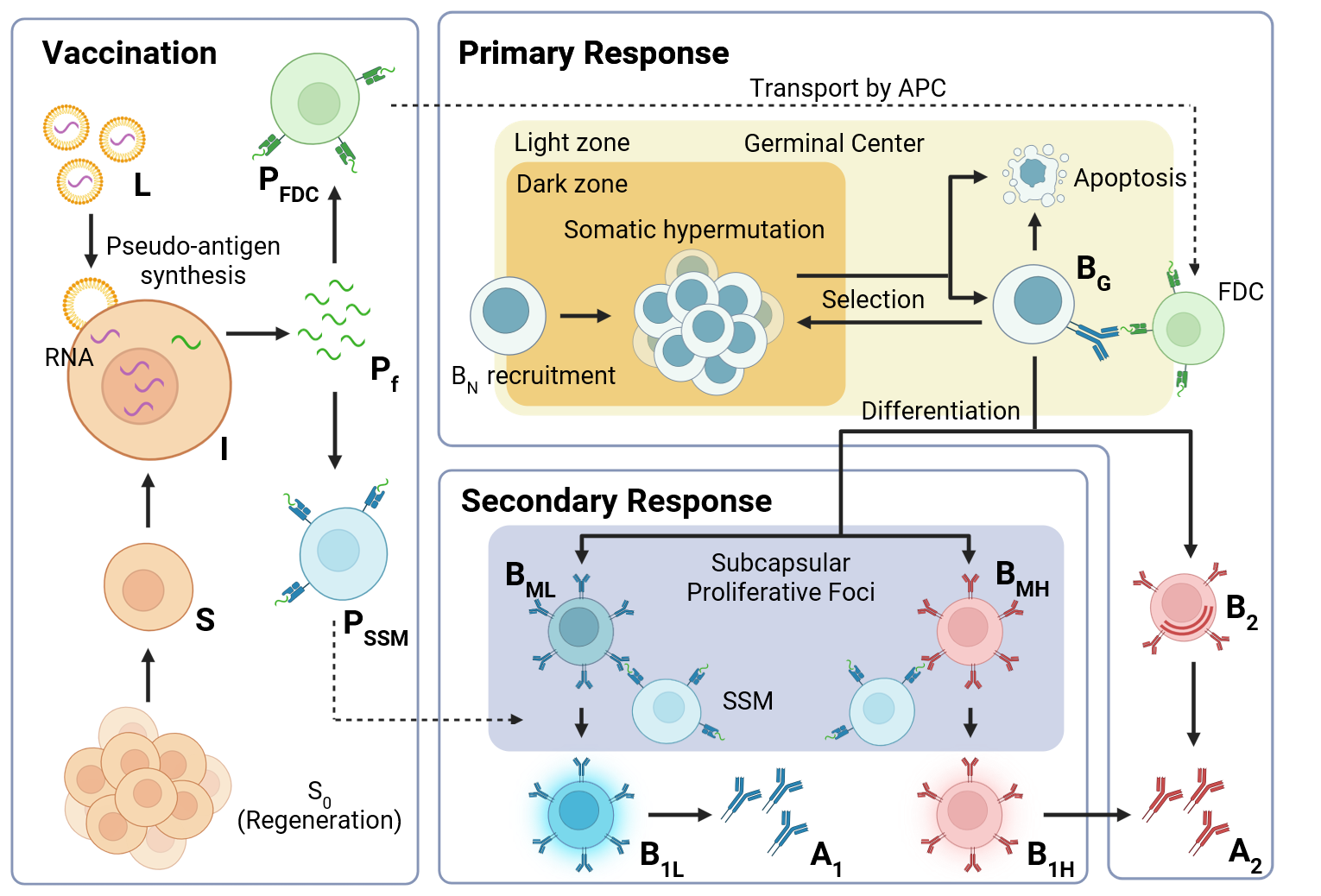}
    \caption{Model schema. Vaccination, followed by production of pseudo-antigen P, induces i) a primary response (activation of GC, affinity maturation (AM) and specification of $B_M$- and $B_2$-cells) and ii) a secondary response (activation of $B_M$ cells, specification of $B_1$-cells). Both contribute to the production of antibodies that help fighting future infection. The model introduced does not comprise details of the transport of antigen and of the AM.}
    \label{fig:1}
\end{figure}

\subsection{Vaccination model}
We present a within-host model of RNA-based vaccination against virus V. The naive host is vaccinated with a V-specific mRNA that is encapsulated in lipid-nano-particles (LNPs). Part of the LNPs is taken up by susceptible cells ($S$) - mostly antibody-presenting-cells (APCs) - which become ‘infected’ cells ($I$). Alternatively, LNPs become degraded over time. Infected cells ($I$) produce pseudo-antigen (P), the protein encoded by the transferred RNA. The antigen triggers an adaptive immune response in the host. In order to describe this response, we extend our model in two steps. In a first step we include the primary response comprising the GC reaction. In a second step, we add the secondary response leading to memory B-cell differentiation. 

\subsection{Primary response model}
Initially, the presence of P leads to activation of na{\"i}ve B-cells ($B_N$). These cells can specify directly into memory B-cells ($B_M$) and short-living plasma cells ($B_1$) or are recruited into developing germinal centers (GCs), where they become germinal center B-cells ($B_G$) \cite{Akkaya2019}. As the $B_N$-based $B_M$- and $B_1$- responses come with antibodies of low affinity against V and on short times scales, we consider the $B_G$-option only. In the GC, $B_G$s underwent affinity maturation (AM), a process that, based on somatic hyper-mutation, improves the binding of their B-cell receptor (BCR) to the antigen. This process comprises expansion, and selection of $B_G$s. Thereby, it requires activity of APCs and of specific T-cell subtypes in the GC. These details are not considered in the model. We assume a continuous increase of the average BCR affinity during V-specific AM. 
$B_G$s leave the GC, thereby specifying either into memory B-cells ($B_M$), or into long-living plasma cells ($B_2$). These cells enrich in subcapsular proliferative foci (SPF) \cite{Moran2018} and the bone marrow \cite{Tellier2018}, respectively. 

\subsection{Secondary response}
In case of prolonged presence of pseudo-antigen or repeated vaccination, $B_M$-cells that have developed in the GCs and moved into SPFs become activated by antigen presented by subcapsular sinus macrophages (SSM) \cite{Moran2018}. Subsequently, they amplify and differentiate into short-living plasma cells ($B_1$) \cite{Seifert2016}. In part, $B_M$-cells re-enter GCs for further maturation. Here, we neglect this option. $B_1$- and $B_2$-cells produce antibodies (A) of different affinity that help fighting infection with V. Waning of the vaccine-induced B-cell responses is analyzed studying the time course of antibody affinity and concentration. 

Parameters of our vaccination and immune response model are derived mainly from studies on RNA-based vaccination against SARS-CoV-2. Thus, in the following, spike protein refers to the SARS-CoV-2 variants of this protein. Most of the related studies focus on the Wuhan variant. Protection against other variants can be simulated e.g. considering the changed antibody-antigen binding affinity \cite{Moulana2023}. Some parameters of the model are taken from animal studies (indicated). A schematic of the entire model i.e. the extended model, is provided in Figure \ref{fig:1}.

\section{Mathematical formulation: Vaccination model}

We describe vaccination as a systemic ‘infection’ with LNPs. Starting at the injection site LNPs distribute passively throughout the body in a few hours \cite{Hassett2024}. We neglect differences between tissues and take blood concentration as the mean. Mainly, APCs uptake LNPs and produce pseudo-antigen \cite{Lindsay2019}. We assume that these antigens distributes passively throughout the body, as well and blood concentration represents the mean concentration of free-floating antigen. However, part of antigen is actively transported by immune cells to lymph nodes leading to high local concentrations within developing germinal centers (GCs) and subcapsular proliferative foci (SPF).

\subsection{LNPs \texorpdfstring{$(L)$}{(L)}}

In the vaccine, RNA is encapsulated in LNPs. Each LNP contains about 5 RNA molecules (1-10, \cite{Bepperling2023}). In the body, the concentration of $L$ decreases due to LNP uptake via endocytosis by susceptible cells ($S$) and LNP degradation.

\begin{equation}
    \frac{\mathrm{d}L}{\mathrm{d}t} = -k_L SL - \frac{L}{\tau_L}.
\end{equation}

Here, $k_L$ is the LNP uptake rate by susceptible cells. We estimate $k_L$ from decay rates of spike-encoding RNA in blood plasma \cite{Kent2024}. Assuming a constant density, $S=S_0$, these decay rates suggest an RNA uptake rate larger than $6.9 \times 10^{-7} \text{mL}/(\text{day}\times \text{cell})$.  Considering 5 RNA molecules per LNP, we assume:  $k_L=2.0\times10^{-7}  \text{mL}/(\text{day}\times \text{cell})$.

We set the lifetime of LNPs in the body equal to the lifetime of encapsulated RNA, $\tau_L=7$ days \cite{Kent2024}. This is a long lifetime for LNPs. Nevertheless, it is supported by experiments detecting Spike-RNA up to 28 days after vaccination \cite{Castruita2023}. 

\subsection{Susceptible cells \texorpdfstring{$(S)$}{(S)}}

LNPs are taken up mostly by APCs including dendritic cells and macrophages (primates, \cite{Lindsay2019, Hassett2024}). Susceptible cells ($S$) that uptake LNPs become ‘infected’ cells ($I$). Accordingly, the density $S$ decreases in the presence of $L$. $S$ decreases also, if susceptible cells capture free-floating antigen ($P_f$) secreted by infected cells, become activated and start traveling to the lymph nodes (see below). $S$ increases if na{\"i}ve APCs travel to the vaccination site. In our model, this process is described as cell-inflow from an external pool that ensures that the initial density of susceptible cells $S_0$ is restored. Thus, changes of $S$ are described as: 
\begin{equation}
    \frac{\mathrm{d}S}{\mathrm{d}t} = -k_S LS -k_{APC}P_f S + k_1(S_0-S).
\end{equation}

Here, $k_S$ is the infection rate of susceptible cells by LNPs. We calculated $k_S$ assuming that one LNP infects one cell ($k_L/k_S =1$ copy/cell). $k_{APC}$ is the activation rate of susceptible cells per captured antigen (see below). Susceptible cells are recruited from a cell pool that has a constant density $S_0$. We estimated it from the density of tissue macrophages in muscle ($10^6$  cells/mL, \cite{Sender2023}).  The rate of recruitment $k_1$ was set to $1/14 \; \text{days}^{-1}$. 

\subsection{Infected cells \texorpdfstring{$(I)$}{(I)}}

Infected cells start synthesis of pseudo-antigen (P, see below) the protein encoded by the transferred RNA. The density of $I$ changes as:

\begin{equation}\label{eq:fullI}
    \frac{\mathrm{d}I}{\mathrm{d}t} = k_S SL - \frac{1}{\tau_I}I.
\end{equation}

Here, $\tau_I$ is the lifetime of infected cells. We assume $\tau_I=4$ days which is the lifetime of circulating dendritic cells \cite{Dalod2014}. This time is much shorter than the lifetime of follicular dendritic cells (FDC) presenting antigen in the GC (see below), which is in mice larger than two weeks \cite{Heesters2013}.

\subsection{Pseudo-antigens P}
Antigen synthesized by infected cells ($I$) cells is released as free-floating protein ($P_f$) into body fluids \cite{Ogata2021}. We assume that susceptible cells ($S$) capture the protein, become activated, move into lymph nodes and relay it to FDCs in the GC or to SSMs in the SPF (for the sake of simplicity in equal amounts). These cells present it, in form of immune complexes $P_{FDC}$ or $P_{SSM}$ to germinal center B-cells ($B_G$) and memory B-cells ($B_M$), respectively. We do not model the antigen transport into lymph nodes (see e.g. mice studies on $B_G$ cells by Heesters et al. \cite{Heesters2013}). The free-floating protein ($P_f$) and the immune complexes ($P_{FDC}$, $P_{SSM}$) changes as:

\begin{equation}\label{eq:fullPf}
    \frac{\mathrm{d}P_f}{\mathrm{d}t} = p_PI -k_pS P_f - \frac{P_f}{\tau_{Pf}},
\end{equation}

\begin{equation}
    \frac{\mathrm{d}P_{FDC}}{\mathrm{d}t} = \frac{k_P}{2} SP_f - \frac{P_{FDC}}{\tau_{FDC}}, \qquad \frac{\mathrm{d}P_{SSM}}{\mathrm{d}t} = \frac{k_P}{2} SP_f - \frac{P_{SSM}}{\tau_{SSM}}.
    \tag{5A,B}
\end{equation}
\setcounter{equation}{5}

The production rate was set to: $p_P=25 \; \text{copies}/(\text{day}\times \text{cell})$  (BHK cells, \cite{Sutton2023}). The lifetime of the free-floating protein $\tau_{Pf}$  was set to 2 days. Under this setting, protein concentration reaches baseline 1-2 weeks after vaccination in agreement with Cognetti et al. \cite{Cognetti2021}. $P_{FDC}$ can survive long times \cite{Heesters2013}. We set its lifetime, $\tau_{FDC}$, to 20 days.  Assuming that it can be released again in small amounts might explain that the Spike protein remains more than 10 weeks detectable in blood \cite{Brogna2023}. There is no report about a similar behavior for $P_{SSM}$. Thus, we assume:  $\tau_{SSM}=\tau_{Pf}$.
We consider degradation to represent the dominant process responsible for loss of free-floating protein. Thus, for the rate of $P_f$ capture by $S$-cells should hold: $k_P S \ll 1/\tau_Pf$.  This relation is ensured for $S \leq S_0$ setting: $k_P=1.0\times10^{-7} \text{mL}/(\text{day}\times \text{cell})$. We calculated $k_{APC}$ assuming that one antigen activates one cell ($k_P/k_{APC} =1$ copy/cell).

\subsection{Injected LPC density}
According to Kent et al. \cite{Kent2024}, the measurable concentration of total RNA peaks in the peripheral blood at day 1 after vaccination at concentrations between $10^5$ and $10^6$ copies/mL. To cope with these findings, we set: $L_0=2.5 \times 10^5$  copies/mL.

\subsection{Second vaccination time}

Our simulations start with the first dose of the vaccine $L_0$ at $t = 0$. Let $t^* > 0$ be a time at which a second dose of the vaccine is administered. Such a subsequent dose of vaccines is simulated as an increase of $L$ by the value $L_0$ at time $t^*$.

\subsection{Analytical solutions}

\begin{figure}[!htb]
    \centering
    \includegraphics[width=\linewidth]{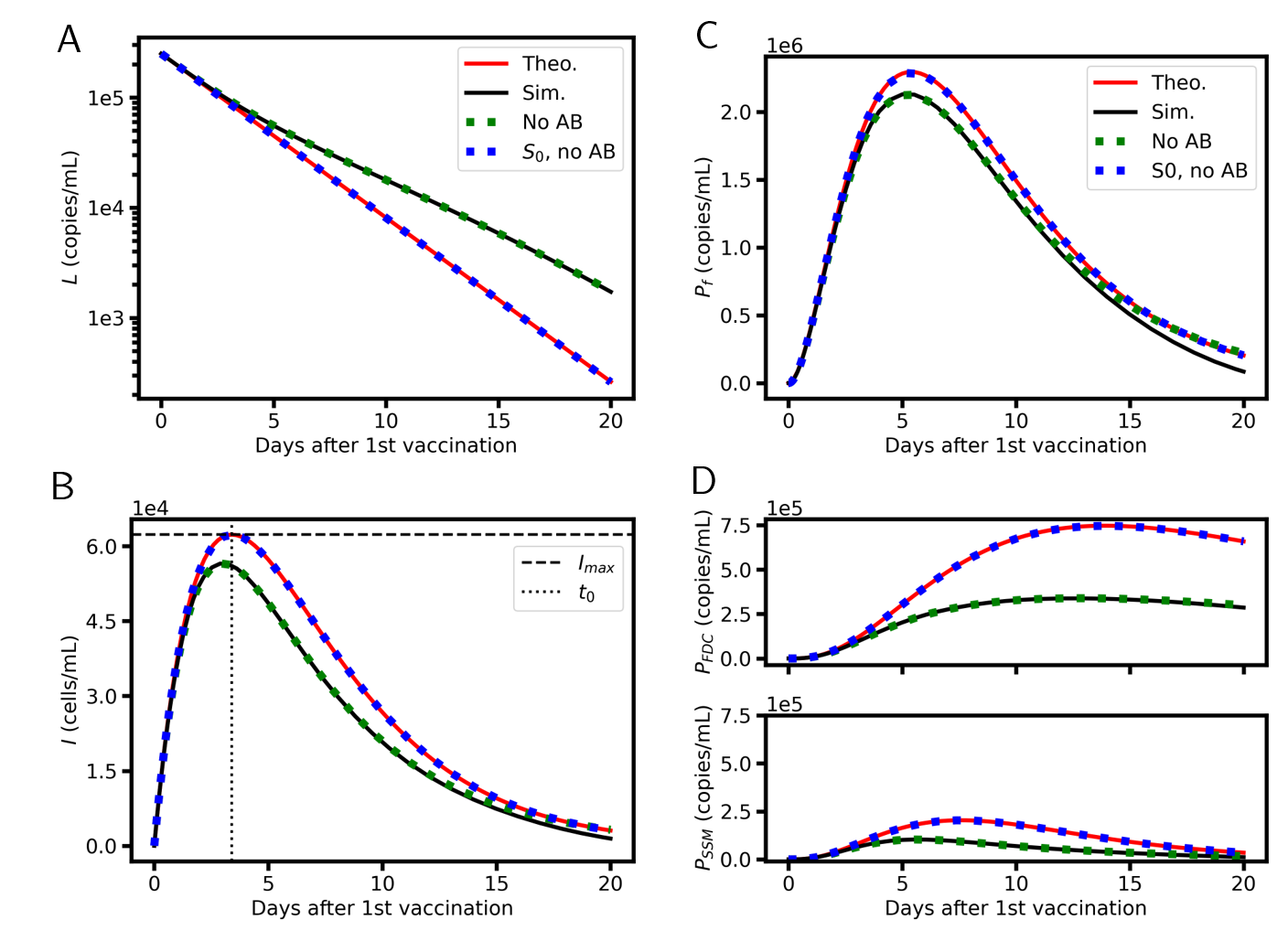}
    \caption{Time course of the concentration of: A) free LNPs (L), B) infected cells (I), C) free-floating protein ($P_f$) and D) interaction complexes ($P_{FDC}$, $P_{SSM}$) after a single vaccination. Shown are analytical (red) and numerical solutions (black lines and symbols). The maximum density of infected cells $I_{max}$ and its time point after vaccination $t_0$ are specified in the Supplement \ref{supp:1} }
    \label{fig:2}
\end{figure}

The vaccination model can be solved analytically if $S$ is kept constant, $S=S_0$ (Fig. \ref{fig:2}). The equations are given in the Supplement \ref{supp:1}. For the LNP concentration ($L$) one obtains an exponential decay. The density of infected cells ($I$) shows a single peak. For the reference model, the peak is reached after about 3 days. The concentrations of free-floating protein ($P_f$) and of interaction complexes ($P_{SSM}$, $P_{FDC}$) peak a few days after. For the reference parameter set, the maximum concentration of $P_f$ is reached after about 5 days in agreement with Ogata et al. \cite{Ogata2021}. According to our settings, $P_{SSM}$ peak at the same time, $P_{FDC}$ after about 14 days.

We compared the analytical with numerical solutions of the extended model described below, and with numerical solutions of the extended model without antibodies (no AB) and without antibodies at constant $S=S_0$ ($S_0$, no AB). We found that: i) newly synthesized antibodies do not affect the behavior before day 13 (see deviation in $P_f$) and ii) variable S has a large impact on $P_{FDC}$ and $P_{SSM}$ (Fig.\ref{fig:2}). 

Thus, in order to analyze the early dynamics of antigen production following RNA-based vaccination the immune response can be neglected. A sophisticated model of the pool of susceptible cells, however, might be required in order to model the B-cell response (depending on $P_{FDC}$ and $P_{SSM}$) quantitatively. Here, we proceed with our simplified model. 

\section{Mathematical formulation: Primary response model}

\subsection{Na{\"i}ve B-cell \texorpdfstring{$(B_N)$}{(BN)}}

Activation of an adaptive immune response requires besides the presence of antigen the presence of na{\"i}ve B-cells ($B_N$) which can become activated. Na{\"i}ve B-cells account for about 50\% of the B-cells in adults \cite{Seifert2016}. Due to their fast turnover, we consider a constant pool of $B_N$ (steady state). As B-cells represent about 10\% of the peripheral blood mononuclear cells \cite{Moore2020}, the $B_N$ density can be estimated by:  $B_N=5 \times 10^4$ cells/mL.

\subsection{Germinal center B-cells \texorpdfstring{$(B_G)$}{(BG)}}
If APCs present antigen within the GC ($P_{FDC}>0$), na{\"i}ve B-cells ($B_N$) become activated and are recruited to become $B_G$s. They start expansion, selection and maturation. Depending on details of the AM process, they differentiate after several days into memory B-cells ($B_M$) or long-living plasma cells ($B_2$).  The GC reaction can take up to 14 weeks \cite{Turner2021}. It includes internalization of antigen ($P_{FDC}$) by $B_G$ cells and thus consumes $P_{FDC}$. We neglect this process in our model. The density of $B_G$ changes as: 

\begin{equation} \label{eq:fullBG}
    \frac{\mathrm{d}B_G}{\mathrm{d}t} = c_N g_{GC} B_N + (c_0 -a_{pD}-\epsilon)B_G - a_{pL}(1-g_{GC})B_G.
\end{equation}

Here, $c_N$ is the maximum recruitment rate of $B_N$ into GCs. With $c_N=10^{-3} \text{day}^{-1}$, the maximum recruitment per day reaches 50 cells/mL. With one GC per ml (see below) this gives 50 cells per GC as suggested by Robert et al. \cite{Robert2024}.

Recruitment is activated by immune complexes ($P_{FDC}$). $B_G$ cells bind $P_{FDC}$ via their BCR. Thereby, they compete with free-floating antibodies A (see below). Thus, the strength of GC activation  $g_{GC}$  depends on the concentrations of immune complexes $P_{FDC}$ and of the antibodies. Details are provided in the Supplement \ref{supp:2}.  

According to Mayer et al. \cite{Mayer2017}, cell amplification ($c_0$) balances the cell apoptosis ($a_{pD}$) in the active GC. Indeed, in mice, a rate of $B_G$- amplification: $c_0=2.0 \; \text{day}^{-1}$ was found \cite{Allen2007}, being in the same range as the apoptosis rate $a_{pD} = 2.0 \;  \text{day}^{-1}$ \cite{Mayer2017}. Thus, we assume: $T= (c_0-a_{pD} )B_G = 0$. The rate $\epsilon$ of cells leaving GC is small. Data by Robert et al. \cite{Robert2024} suggest an upper limit of $0.15 \; \text{day}^{-1}$. We set: $\epsilon = 0.05 \; {day}^{-1}$.  The rate $a_{pL}$ denotes the maximum apoptosis rate in the inactive GC. We set:$a_{pL}= 0.5 \text{day}^{-1}$ ensuing a fast shutdown of the GC if activation $g_{GC}$ decreases. For $T=0$, equation \eqref{eq:fullBG} simplifies to:

\begin{equation} \label{eq:finalBG}
    \frac{\mathrm{d}B_G}{\mathrm{d}t} = c_N g_{GC} B_N - a_{pL}(1-g_{GC})B_G.
    \tag{$6^*$}
\end{equation}
\setcounter{equation}{6}

Thus, for $g_{GC}=1$ (maximum activation) an maximum density $B_G=(c_N/\epsilon)B_N=10^3$ cells/mL can be reached. These cells are confined in the volume of all GCs that supply one mL body fluid. For vaccination into the deltoid muscle, only axillary lymph nodes might be involved in the immune response (20-49 nodes, \cite{Turner2021}). Thus, assuming i) that about 10\% of the 600 lymph nodes in human become activated during vaccination, and ii) 100 GCs per lymph nodes as found in macaques \cite{Margolin2002} an average tissue density of about one V-induced GC per ml can be estimated. Thus, for a GC volume of $2.0\times10^{-5}$ mL \cite{Schemel2023} the local density of $B_G$ can reach: $5.0\times10^7$  cells/mL (see also: Supplement \ref{supp:2}).

\subsection{Affinity maturation}
The activation of the GC reaction and ongoing selection during this reaction depend on the affinity of the BCR of $B_N$- and $B_G$-cells for the antigen, respectively \cite{Wishnie2021}. Affinity is commonly defined as $1/K_D$, where $K_D$ is the dissociation constant of the reaction of interest. For BCR-antigen complex formation, $K_D$ is typically larger than nanomolar \cite{Broketa2023}. We assume a minimal affinity $a_{BG}^0$ that is required for $B_N$ recruitment to the GC. Thus, $B_G$s have an affinity:
\begin{equation}
    a_{BG}=a_{BG}^0 A_{BCR},
\end{equation}
with $A_{BCR} \geq 1$. In the model, only ratios of affinities are relevant (see Supplement \ref{supp:2}). Thus, $a_{BG}^0$ can have arbitrary values. We set $A_{BCR}^{min} =1$ for recruited $B_N$ cells and assume that in course of AM, $A_{BCR}$ and thus, the BCR affinity increases as:
\begin{equation}\label{eq:preABCR}
    \frac{\mathrm{d}A_{BCR}}{\mathrm{d}t} = a_{ff} + w(A_{BCR}^{min}-A_{BCR})= w(A_{BCR}^{max}-A_{BCR})
\end{equation}
The maturation process increases $A_{BCR}$ by a constant value $a_{ff}$ per day. The second term estimates the loss of affinity of the GC cells due to the fact, that cells with low BCR affinity ($A_{BCR}^{min}$) are recruited and cells with higher affinity ($A_{BCR}$) leave the GC.  We set $w$ equal to the fraction of cells that leave the GC, $w = \epsilon= 0.05 \; \text{day}^{-1}$. The maximum $A_{BCR}^{max}=A_{BCR}^{min}+(a_{ff}/w)$ can be calculated for a given $a_{ff}$. It depends on the mutation frequency, which is determined by T-cell help \cite{Cirelli2017}. We set $a_{ff} =1.2$ per day and accordingly $A_{BCR}^{max}=25$. For vanishing V-specific maturation ($a_{ff} =0$), $A_{BCR}$ reduces again to $A_{BCR}^{min}$ (see below).

\subsection{\texorpdfstring{$B_G$}{BG} specification}
$B_G$ cells specify into $B_M$ and $B_2$ cells \cite{Weisel2016}. We consider two types of $B_M$-cells, $B_{ML}$ and $B_{MH}$, with low ($a_{BML}=a_{BG}^0 A_{BCR}^{min}$) and high ($a_{BMH}=a_{BG}^0 A_{BCR}^{max}$) affinity, respectively. All $B_2$-cells have high BCR affinity ($a_{B2}=a_{BG}^0 A_{BCR}^{max}$). The fractions of $B_G$-cells that specify into $B_{ML}$-, $B_{MH}$- and $B_2$-cells is chosen such that the mean value $a_{BG}$ is conserved \cite{Akkaya2019}.

\subsection{Memory B-cells \texorpdfstring{$(B_M)$}{(BM)}}
Initially, the major fraction of amplifying $B_G$-cells specify into $B_{ML}$-cells. With progressing AM, i.e. increasing $A_{BCR}$, more and more $B_{MH}$-cells are specified. The density of $B_{ML}$- and $B_{MH}$-cells evolves as following:
\begin{equation}
    \frac{\mathrm{d}B_{ML}}{\mathrm{d}t} = f \epsilon B_G - \frac{B_{ML}}{\tau_{BM}}, \qquad f = \frac{A_{BCR}^{max}-A_{BCR}}{A_{BCR}^{max}-A_{BCR}^{min}},
\end{equation}
\begin{equation}
    \frac{\mathrm{d}B_{MH}}{\mathrm{d}t} = v(1-f)\epsilon B_G - \frac{B_{MH}}{\tau_{BM}}.
\end{equation}

The lifetime of $B_M$-cells, $\tau_{BM}$, was set to 18 days, as observed for elderly \cite{Macallan2005}. $v$ is the fraction of high affinity $B_G$-cells that specifies into $B_{MH}$-cells, while a fraction of $(1-v)$ specifies into $B_2$- (plasma) cells. In mice, 10 days after immunization, plasma cells nearly exclusively carry high affinity BCRs and represent less than 3 percent of all cells with high affinity BCRs \cite{Phan2006}. Setting: $v=0.9$, similar properties are observed at maximum response around day 21 (single dose, see results).

\subsection{Long-living plasma cells \texorpdfstring{$(B_2)$}{(B2)}}
$B_2$ specification requires fast and repeated cycling in the active GC \cite{Wiggins2020}. This might be enabled by strong T-cell help, which is triggered by particular high BCR affinity. We assume a stochastic nature of the process. In the model, a small fraction $(1-v)$ of high affinity $B_G$-cells becomes $B_2$-cells if the GC matures,

\begin{equation}
    \frac{\mathrm{d}B_2}{\mathrm{d}t} = (1-v)(1-f)\epsilon B_G - \frac{B_2}{\tau_{B2}}.
\end{equation}

The lifetime of $B_2$-cells, $\tau_{B2L}$, was set to $\tau_{BML} = 180$ days (minimum in mice: 90 days \cite{Manz1997}). $B_2$-cells permanently secrete high affinity antibodies $A_2$ (see below). 


\section{Mathematical formulation: Secondary response model}

In a secondary response to antigen, $B_M$-cells located in SPFs become activated, expand and differentiate into $B_1$-cells that produce antibodies capable of neutralizing antigen and labeling infected cells for lysis. $B_M$ activation depends on the antigen affinity of their BCR \cite{Akkaya2019}. We neglect antigen consumption by $B_M$-cell activation.

\subsection{Memory B-cells \texorpdfstring{$(B_M)$}{(BM)}}
If activated, memory B-cells ($B_M$) are capable of fast amplification and differentiation into short-living plasma cells ($B_1$) \cite{Tangye2003}. Activation is triggered by $P_{SSM}$. $B_M$-cells bind the antigen via their BCR. Thereby, they compete with free-floating antibodies. Under such conditions, the density of $B_{ML}$/$B_{MH}$-cells evolves as following:

\begin{equation}
    \frac{\mathrm{d}B_{ML}}{\mathrm{d}t} = f \epsilon B_G +(1-h)c_1g_{BML}B_{ML} - \frac{B_{ML}}{\tau_{BM}},
    \tag{$9^\#$}
\end{equation}
\begin{equation}
    \frac{\mathrm{d}B_{MH}}{\mathrm{d}t} = v(1-f)\epsilon B_G +(1-h)c_1g_{BMH}B_{MH}- \frac{B_{MH}}{\tau_{BM}}.
    \tag{$10^\#$}
\end{equation}
\setcounter{equation}{11}

The strength of $B_M$ activation $g_{BML}$ and $g_{BMH}$ depends on the concentrations of $P_{SSM}$ and of the antibodies $A_1$ and $A_2$. Details are given in the Supplement \ref{supp:2}. The maximum amplification rate for $B_M$-cells, $c_1$, is set to 0.6 $\text{day}^{-1}$ (mice, at maximum stimulation 3 divisions in 5 days on average, \cite{Tangye2003}). The parameter $h$ quantifies the part of amplifying cells that specify into $B_1$-cells. We set $h=0.6$. 

\subsection{Short-living plasma cells \texorpdfstring{$(B_1)$}{(B1)}} 
Short-living plasma cells are derived from activated memory B-cells ($B_M$). Compared to long-living plasma cells ($B_2$), they have a relatively short lifetime of a few days. Their density evolves as:

\begin{equation}
    \frac{\mathrm{d}B_{1L}}{\mathrm{d}t} = h c_1 g_{BM} B_{ML} - \frac{B_{1L}}{\tau_{B1}}
\end{equation}

\begin{equation}
    \frac{\mathrm{d}B_{1H}}{\mathrm{d}t} = h c_1 g_{BM} B_{MH} - \frac{B_{1H}}{\tau_{B1}}
\end{equation}

We set the lifetime of $B_1$-cells to: $\tau_{B1} =5$ days \cite{Khodadadi2019}. $B_{1L}$ ($B_{1H}$) cells permanently secrete low (high) affinity antibodies $A_1$ and $A_2$, respectively (see below). 

\subsection{Antibodies (A)}
Antibodies with low ($A_1$) and high ($A_2$) affinity are produced by $B_{1L}$-plasma cells and by $B_{1H}$- and $B_2$-plasma cells, respectively. As the antibodies have different antigen affinity, we describe them independently. Their turnover can be described as:  

\begin{equation}\label{eq:fullA1}
    \frac{\mathrm{d}A_1}{\mathrm{d}t} = p_A B_{1L} - \frac{A_1}{\tau_A}
\end{equation}

\begin{equation}\label{eq:fullA2}
    \frac{\mathrm{d}A_2}{\mathrm{d}t} = p_A (B_{1H}+B_2) - \frac{A_2}{\tau_A}
\end{equation}

where $\tau_A =60$ days denotes the lifetime of antibodies (30 days \cite{Barnes2021}, 90 days, \cite{Lau2022}) and $p_A=2.0 \; \text{ng}/(\text{day} \times \text{cell})$ is the antibody production rate (0.8 $\text{ng}/(\text{day} \times \text{cell})$ spleen,  3.4 $\text{ng}/(\text{day} \times \text{cell})$ peripheral blood \cite{Bromage2009}).

We assume that following plasma cell differentiation antibody affinity matches BCR affinity of the cell of origin. Nevertheless, association constants with antigen might differ between surface-bound BCR and antibody, e.g. due to steric hindrance \cite{GarcaSnchez2023}. Details are given in the Supplement \ref{supp:2}.

Protection of the host against virus by antibodies A is at least twofold: Antibodies mark infected cells for destruction by antibody-dependent cell-mediated cytotoxicity (ADCC), antibody-dependent cellular phagocytosis (ADCP) or antibody-dependent complement deposition (ADCD) \cite{Izadi2024}. We assume that all antibodies have this labeling capacity. In addition to labeling cells for destruction, antibodies can directly neutralize antigen, which here means can bind Pf and suppress its functionality. Assuming that all antibodies have both capabilities, we modify equations \eqref{eq:fullI}, \eqref{eq:fullPf} and \eqref{eq:fullA1}, \eqref{eq:fullA2} to:

\begin{equation}\label{eq:finalI}
    \frac{\mathrm{d}I}{\mathrm{d}t} = k_S SL - \frac{1}{\tau_I}I-(\beta_{IA1}A_1+\beta_{IA2}A_2)I,
    \tag{$3^\#$}
\end{equation}

\begin{equation}\label{eq:finalPf}
    \frac{\mathrm{d}P_f}{\mathrm{d}t} = p_PI -k_pS P_f - \frac{P_f}{\tau_{Pf}}-(\beta_{PfA1}A_1+\beta_{PfA2}A_2)P_f,
    \tag{$4^\#$}
\end{equation}

\begin{equation}
    \frac{\mathrm{d}A_1}{\mathrm{d}t} = p_A B_{1L} - \frac{A_1}{\tau_A} -(\gamma_{IA1}I+\gamma_{PfA1}P_f)A_1,
    \tag{$14^\#$}
\end{equation}

\begin{equation}
    \frac{\mathrm{d}A_2}{\mathrm{d}t} = p_A (B_{1H}+B_2) - \frac{A_1}{\tau_A}-(\gamma_{IA2}I+\gamma_{PfA2}P_f)A_2.
    \tag{$15^\#$}
\end{equation}
\setcounter{equation}{15}

Here, $\beta_{IA1}$ and $\beta_{IA2}$ are the rates of cell destruction and $\beta_{PfA1}$ and $\beta_{PfA2}$ the rates of antigen neutralization induced by $A_1$ and $A_2$, respectively.  These processes consume antibodies. $\gamma_{IA1}$, $\gamma_{IA2}$, $\gamma_{PfA1}$ and $\gamma_{PfA2}$ are the rates of consumption per infected cell $I$ and pseudo-antigen $P_f$, respectively.  Details are provided in the Supplement \ref{supp:2}. 

\subsection{Antibody-dependent feedback on AM}

In a negative feedback, antibodies control the activity of the GC (see equation \eqref{eq:finalBG}). We assume that they enter the GC and bind the antigen in competition with $B_G$-cell BCRs \cite{Zhang2013}. This also dampens or even eliminates AM towards dominant antigen V \cite{Robert2024}. For $A_{BCR}$ it follows:

\begin{equation}\label{eq:finalABCR}
    \frac{\mathrm{d}A_{BCR}}{\mathrm{d}t} = v(A_{BCR}^{min}-A_{BCR}+(A_{BCR}^{max}-A_{BCR}^{min})g_{GC})
    \tag{$8^\#$}
\end{equation}
\setcounter{equation}{15}

Thus, equation \eqref{eq:preABCR} yields for $g_{GC}=1$ only. For $g_{GC}=0$, $A_{BCR}$ is forced to decrease to $A_{BCR}^{min}$. Consequently, antigen-specific AM is terminated at the time point where $P_f$ is successfully degraded and/or the antibody titer increases. In vivo, GC reaction can be still active in case the AM target switches to subdominant antigen \cite{Finney2018}. We simulate only dominant antigen-specific AM, thus GC lifetime is shorter than seen during GC reaction in response to complex antigen.

Notably, permanent inflow of non-specific BN has been observed in mice and was considered to result in low affinity of BG cells in late GCs \cite{Hgglf2023, deCarvalho2023}. This suggests that due to AM towards non-dominant antigen initial selection towards dominant antigen ($A_{BCR}^{min}=1$) vanishes and values  $A_{BCR}^{min}<1$ could occur. We neglect this effect.


\section{Simulation results}

\subsection{Systems dynamics}

\begin{figure}[!htb]
    \centering
    \includegraphics[width=\linewidth]{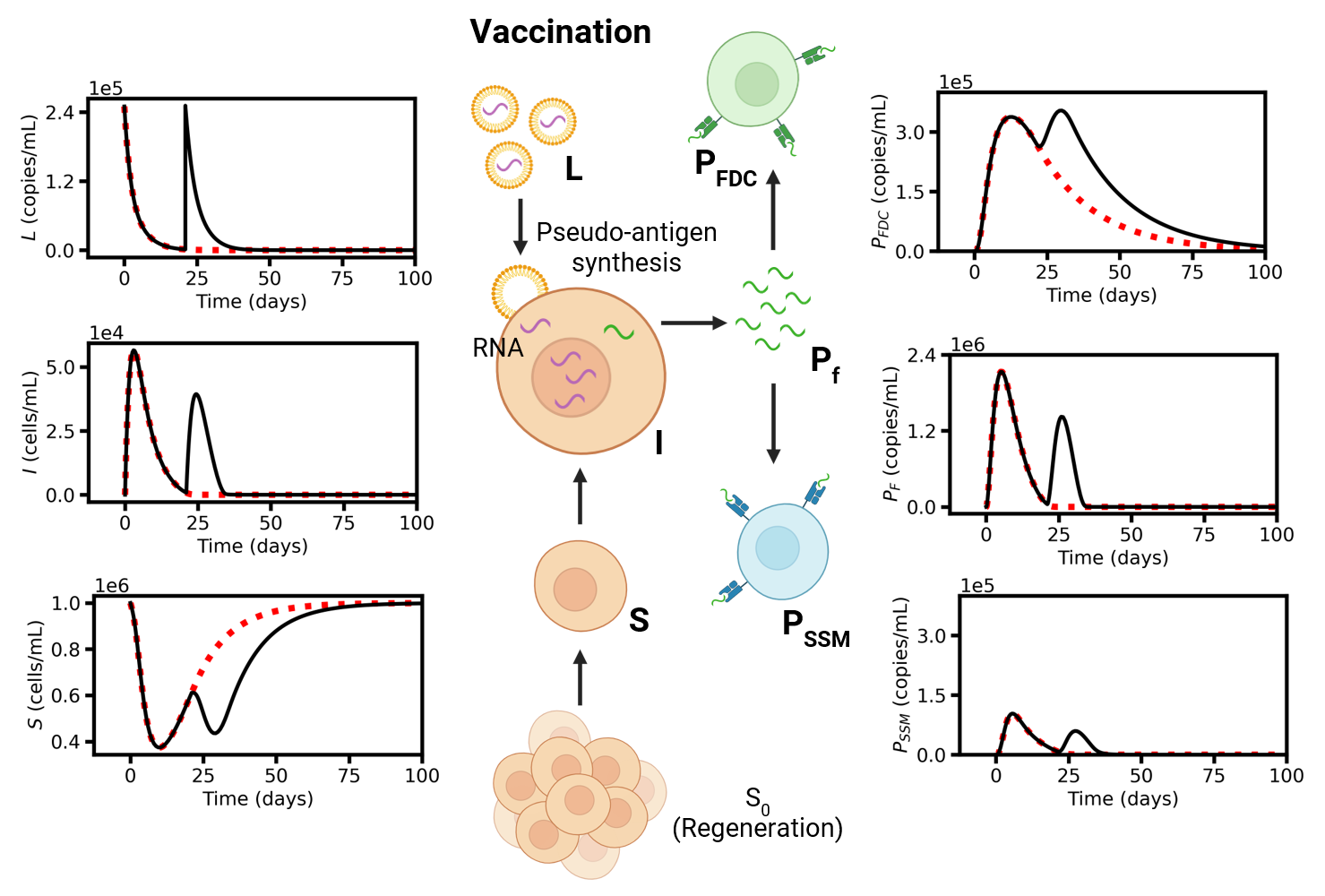}
    \caption{Model dynamics I. Simulation of single (day 0, red dotted line) and double dose vaccination (day 0 and day 21, black solid line) for the reference parameter set. Vaccination model: Concentration of LNPs ($L$), infected cells ($I$), susceptible cells ($S$), free-floating protein ($P_f$), and the two types of immune complexes ($P_{FDC}$, $P_{SSM}$). }
    \label{fig:3}
\end{figure}

\begin{figure}[!htb]
    \centering
    \includegraphics[width=\linewidth]{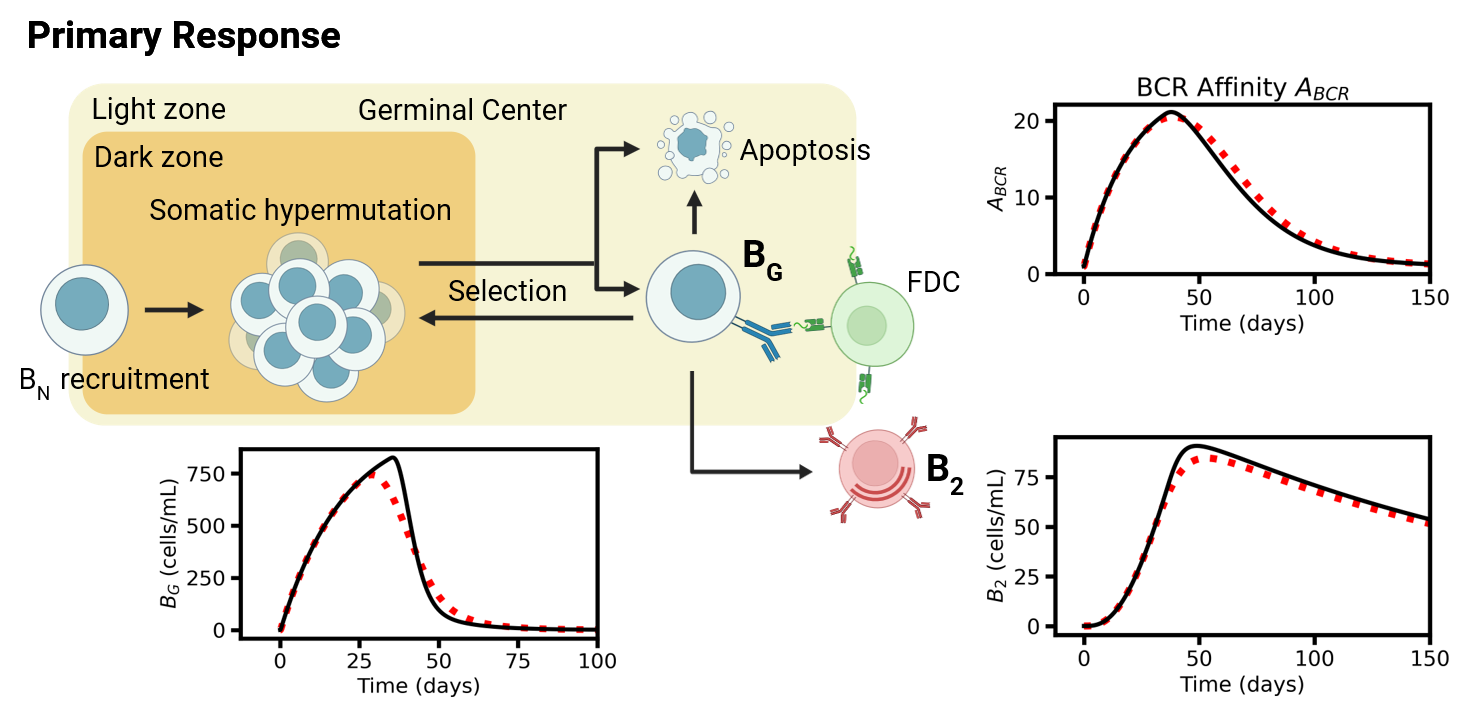}
    \caption{Model dynamics II. Simulation of single and double dose vaccination (see Fig. \ref{fig:3}) for the reference parameter set. Primary response model: Density of GC B-cells ($B_G$), the BCR affinity of $B_G$ cells ($A_{BCR}$) and the density of the long-living plasma cells ($B_2$). }
    \label{fig:4}
\end{figure}

\begin{figure}[!htb]
    \centering
    \includegraphics[width=\linewidth]{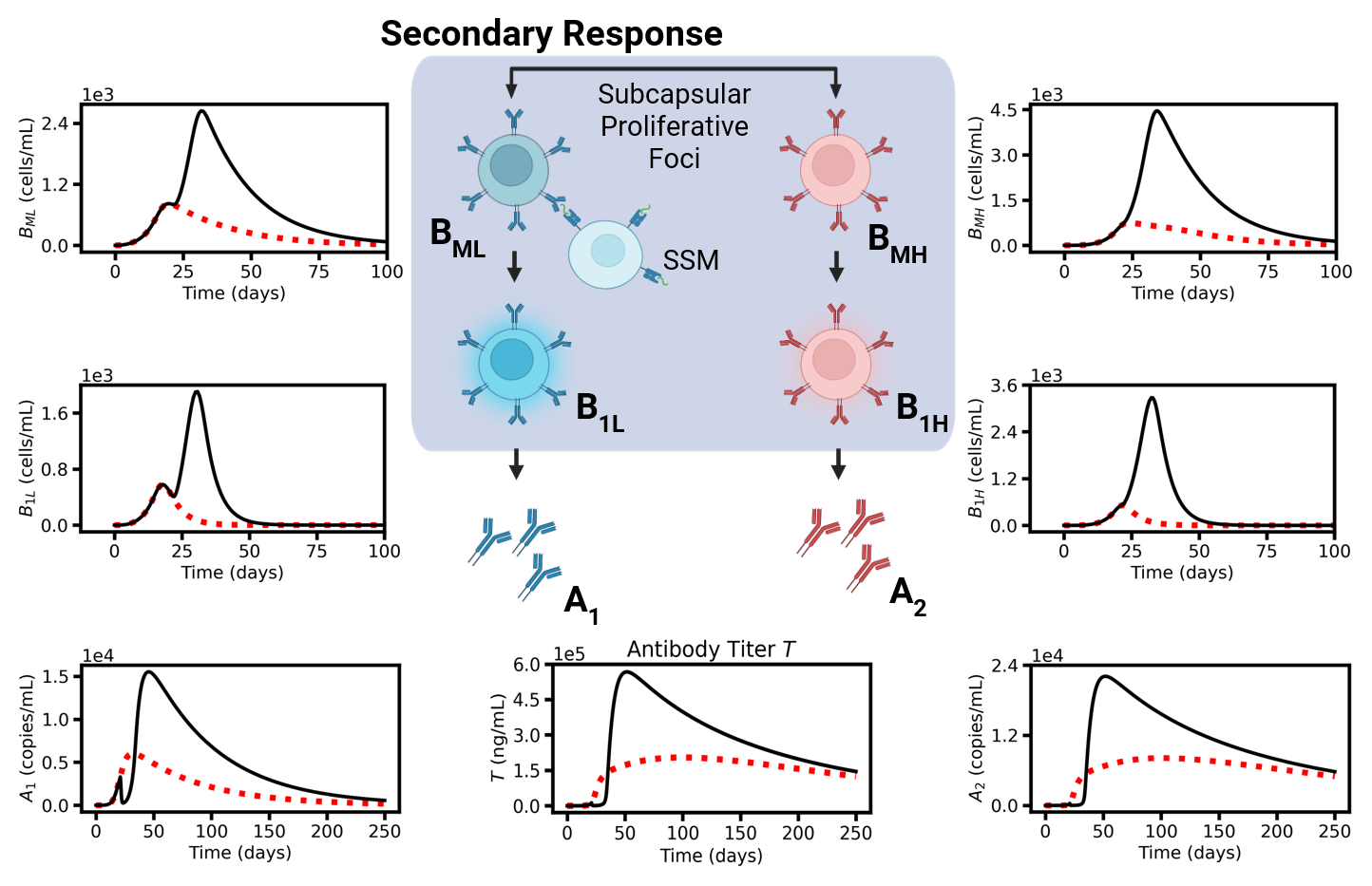}
    \caption{Model dynamics III. Simulation of single and double dose vaccination (see Fig. \ref{fig:3}) for the reference parameter set. Secondary response model: Low and high affinity memory B-cells ($B_{ML}$, $B_{MH}$), low and  high  affinity short-living plasma cells ($B_{1L}$, $B_{1H}$). Protective potential: Low ($A_1$) and high ($A_2$) affinity antibody concentration and the resulting titer T. }
    \label{fig:5}
\end{figure}

The extended model can be solved numerically, only. We started analyzing basic model properties applying the reference parameter set provided in Tables \ref{tab:1a}, \ref{tab:1b} and \ref{tab:1c}. Parameters that were not available from experiments were set such that, the model reproduces properties of the peripheral blood following vaccination against SARS-CoV-2. We compared the response to single and double dose vaccination (Fig. \ref{fig:3}, \ref{fig:4} and \ref{fig:5}).
The first dose stimulates antigen production that peaks a few days later (compare Fig. \ref{fig:2}). At the time point of the second dose, the concentrations of LNPs ($L$) and infected cells ($I$) but also of pseudo-antigen $P_f$ and $P_{SSM}$ have dropped already close to zero (Fig. \ref{fig:3}). The second dose massively increases these concentrations again but has a rather small effect on $P_{FDC}$. Accordingly, the primary response, the GC reaction, is similar for both single and double dose vaccination; in particular, similar numbers of $B_2$ cells are induced (Fig. \ref{fig:4}).  Thus, a second dose does not improve long-term protection very much. 

The second dose controls short-term protection via amplification of the $B_M$-$B_1$ response, i.e. amplification of the secondary response. A single dose vaccination results in a peak concentration of $P_f$ of about 1 to $2 \times 10^6$ copies/mL in agreement with Spike protein peaks (68 pg/mL, \cite{Ogata2021}). This reaction elicits a weak secondary $B_M$-$B_1$ response only (Fig. \ref{fig:5}). Stronger response requires repeated $B_M$ activation and amplification. This is achieved by a second dose, which stimulates production of new antigen. Massive $B_1$ induction results in much higher antibody titers. However, these amplified titers remain for less than 200 days only. Notably, the simulated peak density of $B_M$-cells after the second dose reaches $10^3$-$10^4$ cells/mL and that of antibodies 5 to $10\times10^4$ ng/mL. Both results are in agreement with concentrations found by Goel et al. \cite{Goel2021}.  Moreover, $B_1$ cell densities peak between week 4 and 5 after the first vaccination in agreement with Turner et al. \cite{Turner2021}. 

As shown in Figs. \ref{fig:3}, \ref{fig:4}, \ref{fig:5}, the systems dynamics recapitulate many aspects of RNA-vaccination against SARS-CoV-2. About 2 weeks after the 2nd vaccination, the immune response provides high amounts of antibodies that can help fighting a future infection with the virus. While feedback of these antibodies was negligible for the infection scenario following the first vaccination (first 2 weeks, Fig. \ref{fig:2}), we next asked for its impact on the whole B-cell response.    

\begin{figure}[!htb]
    \centering
    \includegraphics[width=\linewidth]{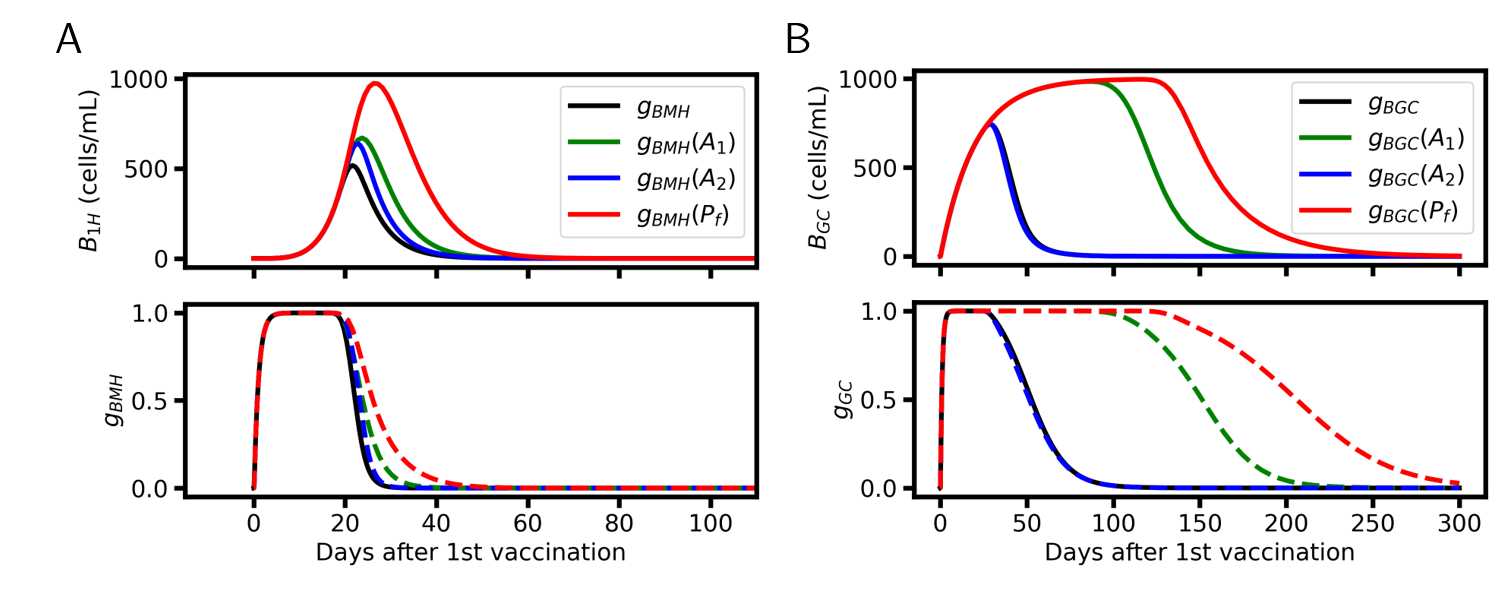}
    \caption{Antibody feedback after a double dose vaccination (reference parameter set).  A) Density of $B_{1H}$-cells (upper row). The emergence of low affinity antibodies is sufficient to stop $B_{1H}$-specification.  $B_{1H}$ peaks at the point, where the activation function $g_{BM}$ (lower row) starts decreasing. Compared is the behavior for full feedback ($g_{BM}$) with that based on $A_1$ ($g_{BM}(A_1)$) or $A_2$ ($g_{BM}(A_2)$) only and the behavior without feedback ($g_{BM}(P_f)$). B) Density of $B_G$-cells (upper row). The shutdown of the GC is triggered by the emergence of $A_2$-antibodies. $B_G$ peaks at the point, where the activation function $g_{GC}$ (lower row) starts decreasing. Compared is the behavior for full feedback ($g_{GC}$) with that based on $A_1$ ($g_{GC}(A_1)$) or $A_2$ ($g_{GC}(A_2)$) only and the behavior without feedback ($g_{GC}(P_f)$). }
    \label{fig:6}
\end{figure}

Detailed analysis revealed that antibody feedback has large impact on plasma cell specification (Fig. \ref{fig:6}). The emergence of antibodies represses $B_M$ activation by competitive $P_{SSM}$-binding with $B_M$-BCRs (Fig. \hyperlink{fig:6}{6A}). Without antibody feedback, the concentration of the $B_{1H}$-maximum would double and arise about 10 days later. To keep it at about 22 days, $A_1$-feedback is sufficient. Shutdown of the GC is controlled by $A_2$-feedback (Fig. \hyperlink{fig:6}{6B}), i.e. by competitive $P_{FDC}$-binding of these antibodies with the $B_G$-BCRs. Thereby, rising $A_2$ concentration limits GC lifetime to about 50 days and thus keeps $B_2$ specification low. Without antibody feedback, the GC would remain activated up to 200 days.

\subsection{Protective potential of vaccination}
The immune response induced by vaccination potentially can protect the individual against infection with virus V. Whether or not protection after some time is given is commonly estimated by the remaining antibody titer. Due to inter-individual differences of the infection, the critical titer required to prevent infection varies as well. 

Here, we used a model of natural infection to calculate the critical titer $T_c$ depending on properties of the virus and the individual immune response. For details, we refer to Supplement \ref{supp:3}. We measure the titer in equivalents of low affinity antibodies $A_1$ in ng/mL. If $T_c\leq0$ ng/mL the infection can be controlled without antibodies. Otherwise, antibodies are required. As both $A_1$ and $A_2$ antibodies contribute to protection, the time-dependent titer acquired by vaccination is given by: 

\begin{equation}
    T = A_1 A_{BCR}^{min}+A_2A_{BCR}^{max} \quad \text{[ng/mL]}
\end{equation}

If $T$ is larger than the critical titer $T_c$, the infection does not spread. Otherwise, it will spread until the immune response on the infection controls it. We analyzed whether a double vaccination as described by our model provides the required level of protection and how long it lasts. 

\begin{figure}[!htb]
    \centering
    \includegraphics[width=\linewidth]{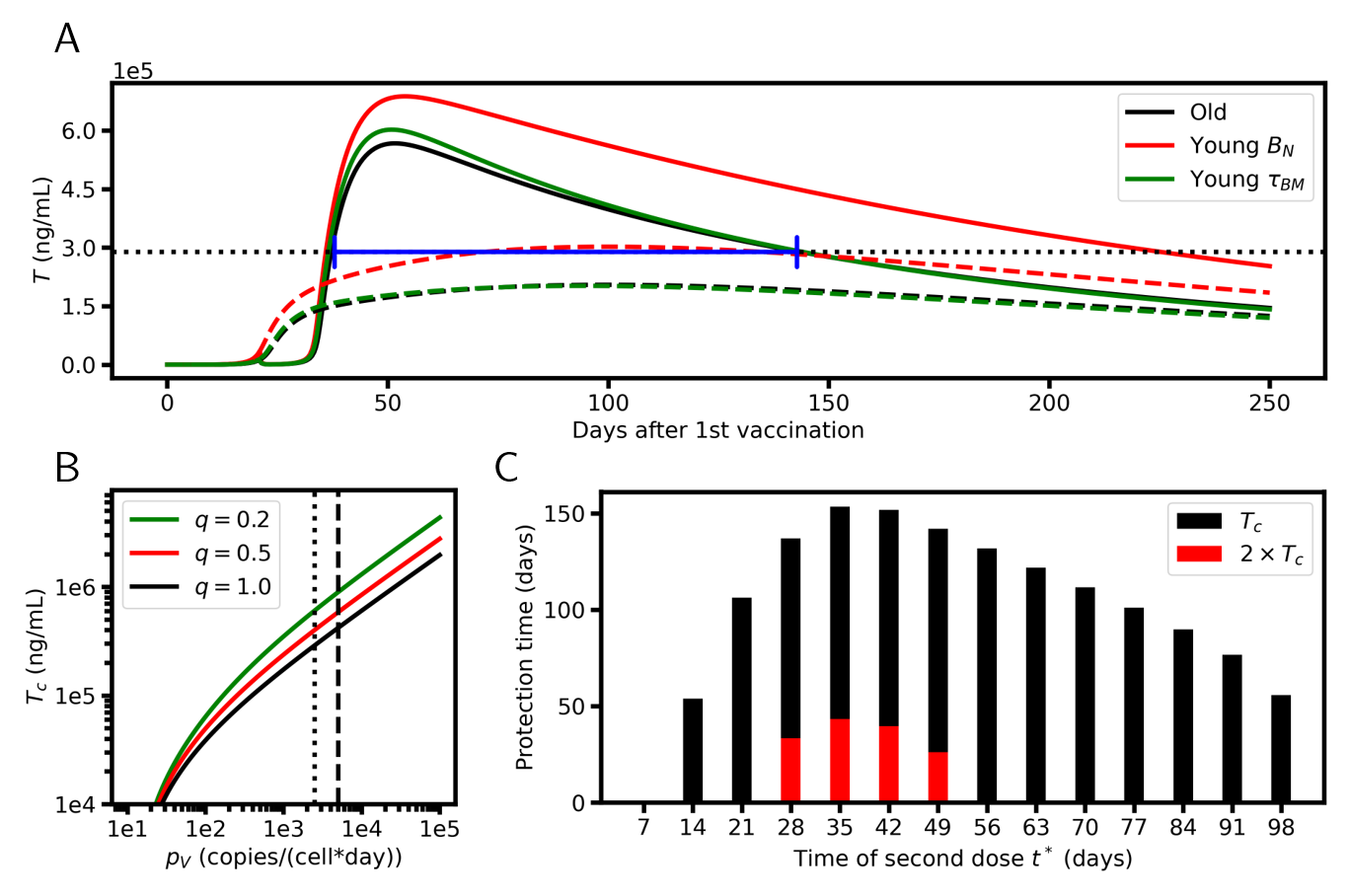}
    \caption{Protection level. A) Simulation results for the titer after a single dose (dashed lines) and a double dose ($t^*=3$ weeks, solid lines) vaccination. The response after a double dose vaccination provides protection against infection (dotted line: $T_c$) for about 15 weeks (blue label). Accounting for properties of young individuals can increase the protection time. B) The critical titer increases with increasing virus replication rate $p_V$ and decreasing antibody-antigen affinities achieved by multiplying them with a factor $q$. C) Protection time vs time to second dose after first dose ($t^*$).  A maximum is seen for $t^*= 5$ weeks.}
    \label{fig:7}
\end{figure}

Fig. \hyperlink{fig:7}{7A} shows simulation results for the titer applying the reference parameter set. The critical titer $T_c$ was calculated as described in the Supplement \ref{supp:3} applying the infection parameter set provided in Tab. \ref{tab:2}. After the first dose, $T_c$ is not reached and no protection is achieved. The second dose ($t^*= 3$ weeks) increases the titer much above $T_c$. This is sufficient to provide protection between day 37 and 143 after first vaccination, i.e. for about 15 weeks.

Notably, we derived part of the reference parameter set from data that are characteristic for blood of elderly. Examples are the lifetime of $B_M$-cells, $\tau_{BM} =18$ days, and the density of na{\"i}ve B-cells, $B_N=5\times10^4$ cells/mL. Changing these parameters to values appropriate to describe blood of younger individuals can increase the protection time. Increasing $\tau_{BM}$ from 18 days to 25 days \cite{Macallan2005} has a minor effect on protection. A strong effect, however, is seen increasing $B_N$ from 0.5 to $1.0 \times 10^5$ cells/mL \cite{Frasca2011}.  The higher $B_N$ density increases the protection time by more than 10 weeks. In this case, even following a single dose the critical titer is reached, although late, i.e. after more than 10 weeks only. 

For mutant virus, the binding strength of antibodies, that have been synthesized following vaccination, to the mutated antigen decreases. Accordingly, higher concentrations of pre-existing antibodies are required in order to avoid future infections. This means that for the synthesized antibodies the critical titer, measured in equivalents of their now lower affinity, increases (Fig. \hyperlink{fig:7}{7B}). Accordingly, the protection time enabled by the vaccination decreases. Mutations can also affect virus replication rates $p_V$. In this case, the critical titer increases with increasing $p_V$ (Fig. \hyperlink{fig:7}{7B})

\subsection{Vaccination timing}
A parameter of the extended model that can be easily controlled to optimize the protection time is vaccination timing $t^*$ (Fig. \hyperlink{fig:7}{7B}). A maximum protection time is observed if the second dose is provided about 5 weeks after the first dose. This is in line with studies by Shioda et al. \cite{Shioda2024}. They found a better protection against SARS-CoV-2 infection for $t^*=26$ to 42 days (Pfizer) and $t^*=33$ to 49 days (Moderna) compared to FDA-recommended $t^*$ of 17-25 days and 24-32 days, respectively. The optimal time for the second vaccination remains stable increasing the critical titer. Thus, vaccination against mutant virus should follow the same timing. 

\begin{figure}[!htb]
    \centering
    \includegraphics[width=\linewidth]{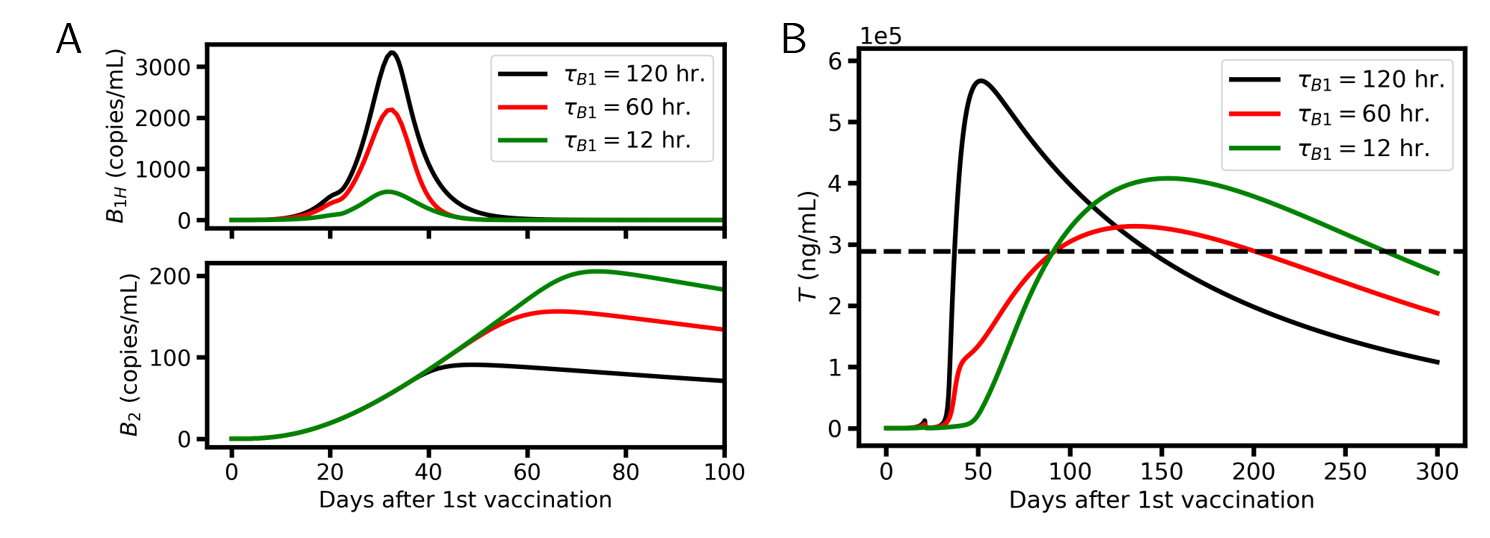}
    \caption{Simulation of $B_{1H}$-cell knock-down by decreasing $B_{1H}$-cell lifetime. A) $B_{1H}$- (upper row) and $B_2$-cell density (lower row). B) Acquired titer. For short $B_{1H}$-cell lifetime, the peak of the titer moves to later times and the protection time increases from 106 to 180 days. }
    \label{fig:8}
\end{figure}

The early occurrence of high affinity $A_2$ antibodies appeared as an essential problem of the vaccination process to induce long-term protection. We have shown (Fig. \ref{fig:6}) that these antibodies shutdown the GC reaction and thus impede the specification of $B_2$-cells. This problem does not occur during natural infection. Here, virus reproduces itself and infects new cells. Accordingly, early emerging high affinity antibodies bind virus and infected cells and do not compete with $B_G$-cells for antigen binding. Thus, the GC reaction remains active until the infection is stopped. In order to support this thesis, we performed $B_{1H}$-cell-knock-down simulations, by shortening $B_{1H}$-cell lifetime (Fig. \ref{fig:8}). In line with our thesis, the number of $B_2$-cells and the protection time increases. However, the peak of the titer moves to later times and protection is reached 10 weeks after the second vaccination only.  

\subsection{Sensitivity analysis}

Our results on the protection time are consistent with the observation of faster waning immunity in elderly \cite{Menni2022, Nanishi2022} and for mutant virus \cite{GarciaBeltran2022}. Next, we study the parameter sensitivity of the protection time and the amplification of antibody titer by a second dose in detail. 

To identify those parameters that contribute most to the variance of these two variables, we performed a variance-based global sensitivity analysis, called the Sobol' method or Sobol' indices \cite{Sobol2001}. This method has been previously used in infectious disease dynamics models, including models for cholera and schistosomiasis \cite{Wu2013}. Sobol' indices rank the importance of each model parameter, depending on how it affects the variance of a model output. An advantage of the Sobol' method is that it both captures the main effects of each parameter (first-order indices), as well as the interaction effects between parameters (total-order indices). Here, we provide the Total Sobol Indices (TSI) as they captures nonlinear effects when two or more parameters are varied at a time.

\begin{figure}[!htb]
    \centering
    \includegraphics[width=\linewidth]{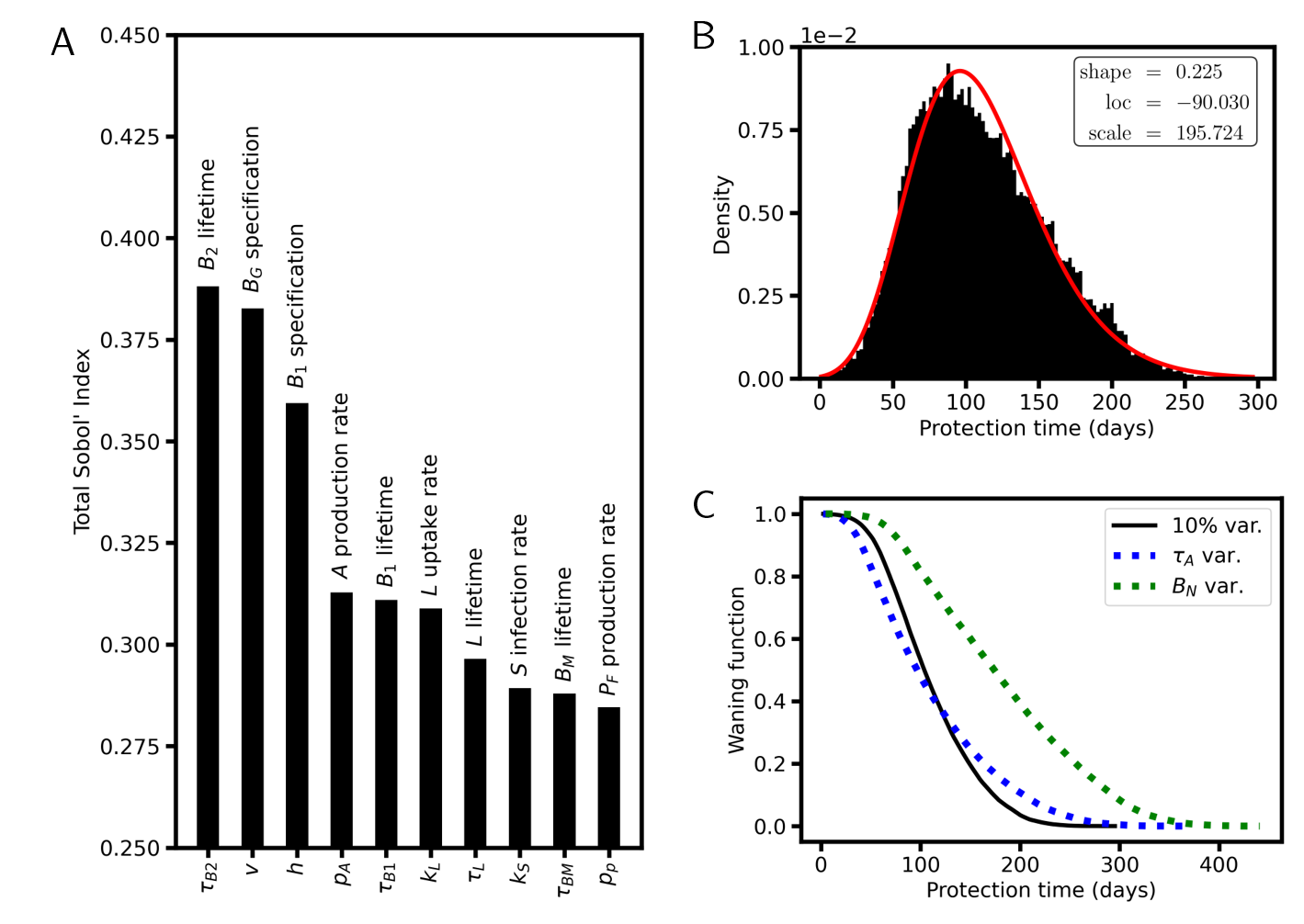}
    \caption{Waning immunity. A) Sensitivity analysis for the protection time including all model parameter assuming 10\% variance for each. The 10 parameters of the extended model with the highest TSI are shown. B) Distribution of the protection time for a population assuming 10\% variation of all model parameters. Fit (red curve): Log-normal distribution (parameters are indicated). C) Waning function for the distribution of the protection time considering 10\% variance and for distributions considering experimental variance for $\tau_A$ (30 to 90 days, \cite{Barnes2021, Lau2022}) and $B_N$ (5 to $15 \times 10^4$ cells/mL). }
    \label{fig:9}
\end{figure}

The three parameters of the extended model that strongest impact protection time are: i) the lifetime of the B2 cells $\tau_{B2}$, ii) the fraction of $B_G$ cells $v$ specifying into $B_M$ cells (respective not into $B_2$), and iii) the fraction of amplifying $B_M$ cells $h$ that specify into $B_{1H}$ (remain not $B_M$ cells) (Fig. \hyperlink{fig:9}{9A}). These are cell intrinsic parameters. It remains open whether they vary depending on the pseudo-antigen encoded by the vaccine, i.e. whether they can be modulated selecting specific antigens. Among the parameter of the infection model, used to calculate the critical titer, the parameters with the highest impact to the protection time are the lifetime of the virus infected cells $\tau_{IV}$ ,  the virus production rate $p_V$ and the virus infection rate $k_V$ (Fig. \ref{fig:S2}). Thus, the potential protection time is most sensitive to the time of virus production, which is typically limited by type 1 IFN response \cite{Paludan2024}, and to virus infection parameters. Together these properties decide whether effective application of RNA-based vaccination is feasible and whether multiple vaccination is required to increase the antibody titer.
Protection could be improved e.g. by modulating parameters linked to plasma cell specification. An example is the parameter $v$, which is the fraction of $B_G$-cells that specify into $B_M$- and not in to $B_2$-cells. Here, improvement can be twofold. Smaller $v$ is linked to longer protection, as the $B_2$-cells are the only antibody producing cells with a lifetime of several months. However, $v$ is linked to the amplification of the protection level after the second dose as well (Fig. \ref{fig:S3}). Larger $v$ results in higher numbers of $B_M$-cells capable of specifying into $B_{1H}$-cells and thus in higher numbers of $B_1$-plasma cells. This enforces amplification of the antibody titer (Fig. \ref{fig:S3}, insert). Thus, for large $v$ a particular high critical titer can potentially be reached. However, the enforced amplification of the titer does reduce the protection time.

Assuming 10\% variance of each parameter, a population of vaccinated individuals shows a distribution of the protection times that peaks around 100 days (Fig. \hyperlink{fig:9}{9B}). The distribution can be fitted by a simple log-normal distribution with a mean protection time of 111 days. One minus its cumulative distribution gives the waning function (WF) of the model (Fig. \hyperlink{fig:9}{9C}).  The decay of the WF becomes broader if experimental variation of $\tau_A$ (30-90 days) or $B_N$ (0.5 to $1.5\times10^5$ cells/mL) is considered. While symmetric variation of the antibody lifetime ($\tau_A$) around 60 days conserves the mean protection time, higher density of na{\"i}ve B-cell ($B_N$), as seen in younger individuals, increases it. Actually, the mean protection time reaches about 189 days, suggesting that the protection time of younger individuals is on average 70\% longer and that of elderly.


\section{Discussion}

A characteristic of RNA-based vaccination against SARS-CoV-2 is fast waning immunity. Here, we introduced a mathematical model of this vaccination method that explains this undesirable characteristic based on an early shutdown of the adaptive immune response by high affinity antibodies. While most of the parameters of the model are derived from vaccination studies against SARS-CoV-2, we expect the principals to hold also if applied to other RNA-virus. 

\subsection{Mechanisms responsible for waning}

An essential part of the model constitutes a negative feedback loop where the antibodies synthesized by the emerging plasma cells shut down the GC reaction as well as the conversion of memory B-cells into plasma cells. Antibody-dependent control of the GC reaction has been suggested already by Zhang et al. \cite{Zhang2013}. How this control mechanism interacts with an ongoing natural infection is largely open. Here, we suggest that ongoing infection consumes newly synthesized antibodies and thus enables enforced $B_2$ specification. In other words, the extensiveness of the infection controls whether a long lasting protection against a secondary infection builds up. In case of RNA-based vaccination, pseudo-antigen producing cells and the antigen itself, in most cases, vanish in short time. We suggest that accordingly a quick increasing concentration of high-affinity antibodies shuts down the GC much too early to enable relevant $B_2$ specification, i.e. long-term protection. Thereby, pre-existing immunity, e.g. cross-reactive memory B-cells that are recalled into GCs, might affect timing \cite{Termote2025}. We neglected these effects in the model.

Vaccination against SARS-CoV-2 can induce long-lasting GC reactions ($>30$ weeks, \cite{Laidlaw2021}) as typically observed for virus infections \cite{Finney2018}. This seems to argue against our thesis. However, AM against the dominant antigen might not proceed over such long periods. Experimental studies documented permanent inflow of BN with low affinity BCR into the GC \cite{deCarvalho2023}. Such behavior has been suggested to originate in competitive AM with so-called ‘dark antigen’ \cite{Finney2018}. The antibody feedback proposed in our model would block AM against dominant antigen after some time and favor dark antigen-specific AM as suggested by experiments \cite{Hgglf2023}. Focusing on dominant antigen specific AM, we did not incorporate these processes in our model. Notably, a recent publication by Mulroney et al. \cite{Mulroney2023}, highlights a possible source of ‘dark antigen’ in case of vaccination against SARS-CoV-2. The authors show, that incorporation of N1-methylpseudouridine into mRNA results in +1 ribosomal frameshifting. Strikingly, +1 frameshifted products from BNT162b2 vaccine mRNA translation occurs after vaccination. As this products reach up to 8\% of the produced protein, competitive AM might indeed be important in case of SARS-CoV-2 vaccination.

Different processes contribute to antibody consumption. Besides direct antigen neutralization, also ADCC consumes antibodies. This process is well studied with respect to its function in SARS-CoV-2 infection \cite{Chen2021}. Its efficacy depends on the targeted cell type. We here assumed that infected cells are mainly APCs. In experiments, these cells show different responses to ADCC. While peripheral dendritic cells are sensitive \cite{Wang2024}, macrophages can escape ADDC \cite{Clayton2021}. Thus, it is currently not clear whether ADCC, while reducing antibody concentration, plays a substantial role in controlling numbers of ‘infected’ cells during vaccination. Similar arguments apply to ADCP and ADCD.

Our model describes the immune response to a single pseudo antigen P, where recruitment of naïve B-cells ($B_N$) to GCs is limited to $c_NB_N$. What happens if vaccines are applied that contain RNA encoding different pseudo antigens $P_1$ and $P_2$? In rodents, RNA-based vaccines combining RNA encoding spike- and nucleocapsid-protein have been investigated \cite{Hajnik2022, Kim2025}. These studies show that the response is additive. This might be a consequence of recruitment of $B_N$s with different BCRs. Thus, in the model one would assume a maximum recruitment $c_{N1}BN+c_{N2}B_N$, with $c_{N1}$ and $c_{N2}$ being the maximum recruitment rates into $P_1$- and $P_2$-specific GCs, respectively.

If the antigens both are V-specific, a combined vaccination provides a better protection against infection with V compared to such against a single antigen. In particular, protection provided by $B_2$-cells secreting antibodies against $P_1$ will combine with that of $B_2$-cells secreting antibodies against $P_2$. Together these antibodies could reach the critical titer and provide a long-term protection. Here, one has to consider that antibodies that do not block virus entry receptors, as e.g. those against nucleocapsid-protein, do not provide neutralization function. Moreover, in mice, spike-nucleocapsid-based vaccines have been found to be associated with high frequency lung pathology \cite{Kim2025}. 

Large differences are observed in the timing of a natural infection as such by SARS-CoV-2, \cite{Owens2024}. According to our results, these differences can be expected to result in large differences in long-term protection against re-infection. Thus, in order to estimate protection times, models of the immune response to natural infection have to consider timing and spreading of the infection carefully. 

\subsection{Waning immunity in cohorts}

We have shown that, variability of immune response in individuals, originating e.g. in different immune cell densities, results in a broad distribution of waning rates in accordance with experimental studies \cite{Menni2022, Nanishi2022}. Unfortunately, biological variances of essential parameters, such as the rates of $B_1$ or $B_2$ specification, are currently unknown. Thus, a quantitative modelling of waning functions is still not feasible. Nevertheless, our studies enable estimations of the importance of individual parameters for long-term protection. Among those showing age-related changes, the size of the compartment of na{\"i}ve B-cells ($B_N$) appears to be important. High numbers improve the protection induced by vaccination. This suggests that comorbid conditions or repeated antigenic exposures that reduce na{\"i}ve B-cell pools may impair vaccine-induced responses. Actually, HIV-infection, which is linked to reduced BN numbers \cite{Wang2024b}, is accompanied by weaker response to SARS-CoV-2 vaccination \cite{Griffin2023}. This holds also for vaccination of HIV-positive individuals against other infections. Whether there is a causal link to reduced $B_N$-cell density remains open. 

We like to emphasize, that our model only explains protection against infection i.e. their probability, but not its severity. Severity of infection might correlate with immune response \cite{Holder2022}, prediction of it from acquired antibody titers alone will fail. Currently, even an estimate of the critical titer required to prevent infection with the original or derived virus strains is challenging. Virus variants, with mutated antibody binding site or changed replication rates modulate the vaccine-induced immunity. Immunization against the Wuhan variant of SARS-CoV-2 was very efficient. The Omicron variant showed high escape from vaccination-induced immunization due to mutations in the receptor-binding domain of the Spike protein. Changes of the BCR affinity distribution following SARS-CoV-2 evolution have been quantified \cite{Moulana2023}.

Our model can consider such changes by changing antibody-BCR interaction strength. More sophisticated modeling strategies that integrate binding site variability have been introduced recently \cite{Robert2024}. However, for quantitative modeling of infection scenarios in parallel sophisticated spatio-temporal models of infection spreading are required. This process can vary even between closely related virus strains e.g. due to variable tropism \cite{Yang2025}.

Mathematical modeling of vaccination scenarios can suggest optimal timing of the vaccination process and necessary modifications for virus variants. However, quantitative recommendations require validation against clinical/serological datasets. In addition, mathematical modelling can support understanding of the general kinetics of the adaptive response and help to avoid pitfalls of new approaches.     


\vspace{6pt} 





\authorcontributions{Conceptualization, J.G. and T.K.; methodology, J.G.; software, J.M.; formal analysis and investigation, J.G.; data curation, J.M.; writing---original draft preparation, J.G.; writing---review and editing, J.G., J.M.; visualization, J.M. All authors have read and agreed to the published version of the manuscript.}

\dataavailability{The source code for the model is available on \url{https://github.com/jarmsmagalang/immu_buildup_waning}} 
\acknowledgments{This work was supported by the UniBE Short Travel Grants for (Post)Docs 2024. Figures \ref{fig:1}, \ref{fig:3}, \ref{fig:4}, \ref{fig:5} have been created with \url{https://BioRender.com}.}

\conflictsofinterest{The authors declare no conflicts of interest.} 



%

\appendixtitles{yes} 
\appendixstart
\appendix
\section[\appendixname~\thesection]{Analytical solutions for the vaccination model}
\label{supp:1}
For $S=S_0$, the concentration of LNPs is given by:

\begin{equation}
    L = L_0 \exp \left[ - \left( k_L S_0 + \frac{1}{\tau_L} \right)t \right].
\end{equation}

Setting $k_L S_0 +(1/\tau_L) = 1/\tau_{LC}$, one obtains the density of infected cells:

\begin{equation}
    I = I_0 \left[ \exp\left(- \frac{t}{\tau_I} \right) - \exp \left(-\frac{t}{\tau_{LC}} \right) \right], \quad \text{with } I_0 = \frac{k_LS_0L_0\tau_I}{(\tau_I/\tau_{LC})-1}. 
\end{equation}

The maximum density of infected cells is given by

\begin{equation}
    I_{max} = k_L S_0L_0\tau_I x^{-\frac{x}{x-1}}, \quad \text{with } x = \frac{\tau_I}{\tau_{LC}}>1.
\end{equation}

It is reached at $t_0 = (\tau_I \ln x)/(x-1)$. Setting $k_P S_0 + (1/\tau_{Pf}) = 1/\tau_{PC}$, one obtains for the concentration of free-floating protein:

\begin{equation}
\begin{split}
    P_f = & P_0 \Biggl[ \frac{\tau_I}{\tau_I-\tau_{PC}}\left(\exp \left(-\frac{t}{\tau_I} \right) - \exp \left(- \frac{t}{\tau_{PC}} \right) \right) \\
    & -  \frac{\tau_{LC}}{\tau_{LC}-\tau_{PC}}\left(\exp \left(-\frac{t}{\tau_{LC}} \right) - \exp \left(- \frac{t}{\tau_{PC}} \right) \right) \Biggr], \quad \text{with } P_0 = p_PI_0\tau_{PC},
\end{split}
\end{equation}

and for the immune complex $P_{FDC}$ (similarly for $P_{SSM}$ but with $\tau_{SSM}$):

\begin{equation}
    \begin{split}
        P_{FDC} = & \frac{k_P}{2}S_0 P_0 \tau_{FDC} \Biggl[ \frac{\tau_I^2}{(\tau_I-\tau_{PC})(\tau_I-\tau_{FDC})} \left( \exp \left( -\frac{t}{\tau_I}\right) - \exp \left( -\frac{t}{\tau_{FDC}}\right) \right) \\
        &-\frac{\tau_{LC}^2}{(\tau_{LC}-\tau_{PC})(\tau_{LC}-\tau_{FDC})} \left( \exp \left( -\frac{t}{\tau_{LC}}\right) - \exp \left( -\frac{t}{\tau_{FDC}}\right) \right) \\
        &+\frac{\tau_{PC}^2(\tau_I-\tau_{LC})}{(\tau_{PC}-\tau_{I})(\tau_{PC}-\tau_{LC})(\tau_{PC}-\tau_{FDC})} \left( \exp \left( -\frac{t}{\tau_{PC}}\right) - \exp \left( -\frac{t}{\tau_{FDC}}\right) \right) \Biggr]
    \end{split}
\end{equation}

\begin{figure}[!htb]
    \centering
    \includegraphics[width=\linewidth]{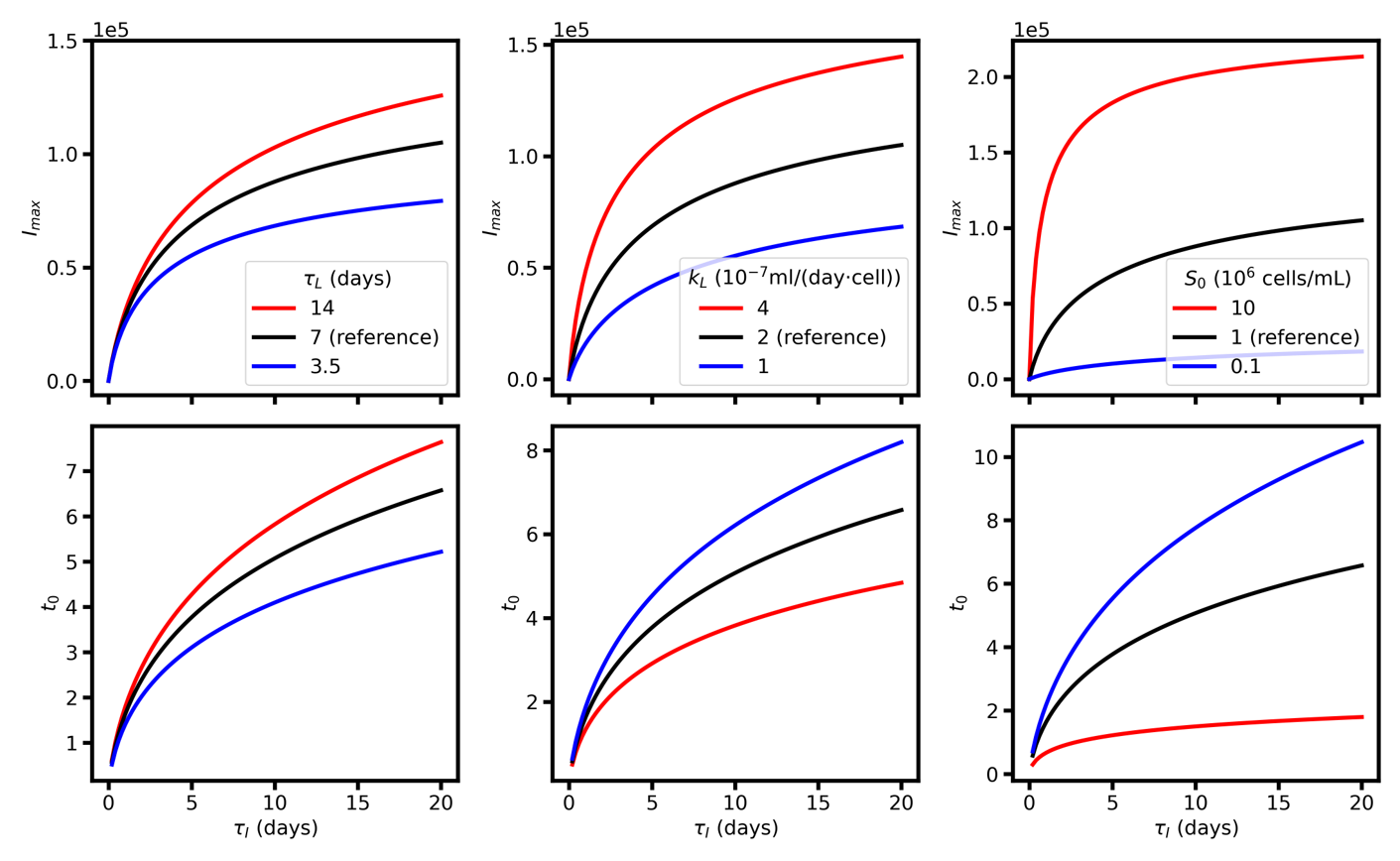}
    \caption{Maximum density of infected cells $I_{max}$ following vaccination.  For larger life-time of the infected cells, $\tau_I$, the maximum occurs later and reaches higher densities. Similar changes are observed for larger $\tau_L$. Increasing $k_L$ and $S_0$, however, increases the maximum cell density but shorten the time to reach it.}
    \label{fig:s1}
\end{figure}

\section[\appendixname~\thesection]{Details on antibody feedback}
\label{supp:2}
Activation of the GC depends on the concentrations of immune complex $P_{FDC}$ and of the antibodies $A_1$ and $A_2$. $B_G$ cells bind the antigen $P_{FDC}$ via their BCR. Thereby, they compete with free-floating antibodies. Assuming a competitive binding of antigen by BCRs and antibodies, classical enzyme kinetics gives for the concentration of the antigen-BCR complex:

\begin{equation}
    [P_{FDC}B] = \frac{P_{FDC,GC}}{1+ \frac{K_B}{B} + \frac{K_B}{K_A} \frac{A}{B}},
\end{equation}

where $P_{FDC,GC}$ is the total concentration of $P_{FDC}$ in the GC, and $K_B$, $K_A$ the dissociation constants of BCR- and A- binding, respectively.  The BCR density is given by: $B=nB_{G,GC}$. Here, $n$ is the number of BCR per $B_G$ cell [copies/cell] and $B_{G,GC}$ the concentration of $B_G$ cells in the GC. The concentration $[P_{FDC}B]$ divided by $B_{G,GC}$  is the number of complexes per BG cell [copies/cell]:

\begin{equation}\label{eq:app_pfdcb}
    \frac{[P_{FDC}B]}{B_{G,GC}} \approx \frac{P_{FDC,GC}}{B_{G,GC} + \frac{K_B}{K_A} \frac{A}{n}}.
\end{equation}

We assume that the activation $g_{GC}$ of the germinal center requires that a sufficient number $1/C$ [copies/cell] of BCRs are bound by antigen. 

\begin{equation}\label{eq:app_ggc}
    g_{GC} = 1-\exp \left(- \frac{[P_{FDC}B]}{B_{G,GC}}C \right)
\end{equation}

Typically, $10-20$ bound BCR are required to activate a signal in B-cells. We set: $C=0.1$. We estimate the parameters of equations \eqref{eq:app_pfdcb} and \eqref{eq:app_ggc} as following: $P_{(FDC,GC)}$ can be calculated from $P_{FDC}$ considering that the complexes are concentrated in GCs (1 GC/mL), leading to $P_{(FDC,GC)}=P_{FDC} \; \text{mL} / 1 \, V_{GC}$ . With a GC-volume of $V_{GC} =2\times10^{-5}$ mL \cite{Schemel2023}, it follows: $P_{(FDC,GC)}=5\times10^4 \; P_{FDC}$. The same factor applies to the concentration of $B_G$s: $B_{G,GC}= 5 \times 10^4 \; B_G$. Calculating A in ng/mL, it follows:

\begin{equation}
    \frac{[P_{FDC}B]}{B_{G,GC}} \approx \frac{P_{FDC,GC}}{B_{G,GC} + \frac{K_B}{K_A} x\frac{A}{n}},
\end{equation}
where $x$ gives the number of antibodies per ng [copies/ng]. The constant $r=(xK_B)/(n K_A)$ [cells/ng] is antibody-specific. For an antibody with the same dissociation constant as the BCR follows: $r=x/n$. For $x=4\times10^9$ copies/ng and $n=10^5$ copies/cell it follows:  $r=4\times10^4$ cells/ng. Thus, activation $g_{GC}$ is given by:

\begin{equation}\label{eq:gGC}
    g_{GC} = 1- \exp\left(- \frac{(5 \times 10^4) P_{FDC}}{(5 \times 10^4) B_G + (4 \times 10^4) \frac{cells}{ng} A} C \right) \approx 1 - \exp \left( - \frac{P_{FDC}}{B_G + \frac{cells}{ng}A} C\right)
\end{equation}

In case of $B_M$ activation, these cells are concentrated in subcapsular proliferative foci (SPF) and interact with subcapsular sinus macrophages (SSM) that present interaction complexes $P_{SSM}$ \cite{Moran2018}. We assume that SPF have to confine antigen and B-cells similar to the GC. Thus, the concentration of antigen and the density of B-cells are similarly increased in this structure. It follows:  

\begin{equation}\label{eq:gBM}
    g_{BM} = 1- \exp\left(- \frac{(5 \times 10^4) P_{SSM}}{(5 \times 10^4) B_M + (4 \times 10^4) \frac{cells}{ng} A} \right) \approx 1 - \exp \left( - \frac{P_{SSM}}{B_M + \frac{cells}{ng}A} C\right)
\end{equation}

Thus, activation properties are similar to the GC, but depend on the short-living interaction complex $P_{SSM}$.

Equations \eqref{eq:gGC} and \eqref{eq:gBM} describe activation functions for binding competition for antigen between BCRs and antibodies with the same dissociation constants. To consider different affinities of both we assume:

\begin{equation}
    A_{GC} = \frac{A_1 A_{BCR}^{min}+ A_2 A_{BCR}^{max}}{A_{BCR}}, \quad \text{for }g_{GC}
\end{equation}

\begin{equation}
    A_{BML} = \frac{A_1 A_{BCR}^{min}+ A_2 A_{BCR}^{max}}{A_{BCR}^{min}}, \quad \text{for }g_{BML}
\end{equation}

\begin{equation}
    A_{BMH} = \frac{A_1 A_{BCR}^{min}+ A_2 A_{BCR}^{max}}{A_{BCR}^{max}}, \quad \text{for }g_{BMH}.
\end{equation}

The rates of cell destruction, $\beta_{IA1}$ and $\beta_{IA2}$, and the rates of antigen neutralization, $\beta_{PfA1}$ and $\beta_{PfA2}$, depend on the affinity of the antibodies as well. We assume:

\begin{equation} \label{eq:appbetaIA}
    \beta_{IA1} = \beta_{IA}A_{BCR}^{min} \quad \text{and} \quad \beta_{IA2} = \beta_{IA}A_{BCR}^{max},
\end{equation}

\begin{equation}  \label{eq:appbetaPfA}
    \beta_{PfA1} = \beta_{PfA}A_{BCR}^{min} \quad \text{and} \quad \beta_{PfA2} = \beta_{PfA}A_{BCR}^{max}.
\end{equation}

Values of $\beta_{IA}$ and $\beta_{PfA}$ are provided in Tab. \ref{tab:1c}. Similarly, we assume for the antibody consumption rates $\gamma_{IA1}$, $\gamma_{IA2}$, $\gamma_{PfA1}$ and $\gamma_{PfA2}$:

\begin{equation}
    \gamma_{IA1} = \gamma_{IA}A_{BCR}^{min} \quad \text{and} \quad \gamma_{IA2} = \gamma_{IA}A_{BCR}^{max},
\end{equation}

\begin{equation}
    \gamma_{PfA1} = \gamma_{PfA}A_{BCR}^{min} \quad \text{and} \quad \gamma_{PfA2} = \gamma_{PfA}A_{BCR}^{max}.
\end{equation}

Values of $\gamma_{IA}$ and $\gamma_{PfA}$ are provided in Tab. \ref{tab:1c}.

We set $\beta_{PfA}= \beta_{IA}  = 5\times10^{-5} \text{mL}/(\text{day} \times \text{ng})$. Thus, at the maximum concentration of antibodies reached in our simulations (about $6\times10^4$ ng/mL) the antibody-depending terms in equations \eqref{eq:finalI} and equation \eqref{eq:finalPf} reach (-3/($\text{day}\times I$)) and (-3/$\text{day}\times P_f$)), respectively. This means that all infected cells become destroyed and all free protein neutralized within 8h. The amount of antibody required for these processes and thus the parameters $\gamma_{IA}$ and $\gamma_{PfA}$ were estimated from experiments.

\subsection[\appendixname~\thesubsection]{ADCC experiment}

Infected cells (target cells, initial density $I_0$) and neutrophils are mixed with antibodies (initial concentration $A_0$). Infected cells fluoresce due to an integrated reporter gene. The total fluorescence of the population decreases due to ADCC by neutrophils. (time scale: 12 hours). Assuming a constant neutrophil density, we calculate $I$ and $A$ according to:

\begin{equation}
    \frac{\mathrm{d}I}{\mathrm{d}t} = \beta_{IA} A I
\end{equation}

\begin{equation}
    \frac{\mathrm{d}A}{\mathrm{d}t} = \gamma_{IA} A I
\end{equation}

with the rate of cell destruction by ADCC , $\beta_{IA}$ [mL/($\text{$\mu$g} \times \text{day}$)], and the rate of antibody consumption  $\gamma_{IA}$ [mL/($\text{cell} \times \text{day}$)]. 

Here, $n=\gamma_{IA}/\beta_{IA}$ [$\mu$g/cell] is the amount of antibody required per cell. The analytical solution for the remaining infected cells ($I$) is given by: 

\begin{equation}
    I = I_0 \frac{A_0 - nI_0}{A_0\exp \left( t \beta_{IA} (A_0-nI_0)\right) - nI_0}.
\end{equation}

For $A_0<n I_0$, one observes for long times

\begin{equation}
    I = I_0 (1- \frac{A_0}{nI_0}),
\end{equation}

otherwise, $I = 0$.

In Alpert et al. \cite{Alpert2012}, they show that for HIV infection, an initial antibody concentration of $A_0=100$ $\mu$g/ml and an initial density of infected cells of $I_0=5 \times 10^4$ cells/mL, allows  20\% of the cells to survive, giving: $n=2.5$ ng/cell.  

\subsection[\appendixname~\thesubsection]{Neutralization experiment}

Pseudo-virus (initial concentration $P_{f0}$) is mixed with antibodies (initial concentration $A_0$). The remaining capability of the mixture to infect cells that fluoresce due to an integrated reporter gene is measured (time scale: 10 minutes).

\begin{equation}
    \frac{\mathrm{d}P_f}{\mathrm{d}t} = \beta_{PfA} A P_f,
\end{equation}

\begin{equation}
    \frac{\mathrm{d}A}{\mathrm{d}t} = \gamma_{PfA} A P_f,
\end{equation}

with the rate of cell neutralization by ADCC, $\beta_{PfA}$ [mL/($\text{ng} \times \text{day}$)], and the rate of antibody consumption, $\gamma_{Pf A}$ [ml/($\text{copy} \times \text{day}$)]. 

Here, $m=\gamma_{PfA}/\beta_{PfA}$ [ng/copy] is the amount of antibody required per antigen.  
The analytical solution for the remaining infected cells I is given by: 

\begin{equation}
    P_f = P_{f0} \frac{A_0 - mP_{f0}}{A_0\exp \left( t \beta_{PfA} (A_0-mP_{f0})\right) - mP_{f0}}.
\end{equation}

For $A_0<m P_{f0}$, one observes for long times

\begin{equation}
    P_f = P_{f0} (1- \frac{A_0}{mP_{f0}}),
\end{equation}

otherwise, $P_f = 0$.

In Chen et al. \cite{Chen2025}, they show that the total neutralization is reached by 10\% of the serum antibody level ($A_0= 1 $ $\mu$g/ml). In their assays they use: $P_{f0}= 500$ TCID50.  Assuming TCID50 $= 4\times10^3$ copies/mL \cite{Bordi2020} this gives: $P_{f0}= 2 \times 10^6$ copies/mL and:  $m= 5\times 10^{-4}$ ng/copy.

Using the values of $n$ and $m$, the parameters $\gamma_{IA}$ and $\gamma_{Pf A}$ were calculated from $\beta_{IA}$ and $\beta_{PfA}$.

\section[\appendixname~\thesection]{Virus infection model}\label{supp:3}

We assume that virus V infects a specific tissue only. The tropism of V is thereby determined by the level of expression of virus entry receptors \cite{Compans2005}. In the following, we model an infection of the respiratory tract. We assume initial cell-free infection. Subsequent spreading depends on the site of virus release. While apical release of virus into lumen is typically associated with local infection, basolateral release can result in systemic infections. We assume virus release into the lumen. Nevertheless, infected immune cells, as macrophages, can contribute to spreading \cite{Wang2024c}.  We neglect this contribution.

\subsection[\appendixname~\thesubsection]{Susceptible cells}

Spreading via cell-cell contacts limits the infection rate \cite{Raach2023}. For a non-spatial population model, we assume:  

\begin{equation}
    \frac{\mathrm{d}S}{\mathrm{d}t} = -k_{VI}\left(1-\frac{I}{I_{max}} \right) SV + k_2(S_0-S).
\end{equation}

The virus infection rate $k_{VI}$ depends on the virus variant. We set: $k_{VI}=10^{-7}$  mL/($\text{day} \times \text{copy}$). This value is in the range that have been assumed for corona viruses \cite{Kim2021}. The maximum density of infected cells typically reaches 10\% of $S_0$ \cite{Ravindra2021}. Regeneration, depends on lung epithelial cell proliferation. 

\subsection[\appendixname~\thesubsection]{Infected cells}

Virus infected cells ($I$) produce virus, until they undergo lysis or become subject of phagocytosis \cite{Nainu2017}:

\begin{equation}
    \frac{\mathrm{d}I}{\mathrm{d}t} = k_{VI}\left(1-\frac{1}{I_{max}}I \right) SV- \frac{1}{\tau_I}I-(\beta_{IA1}A_1+\beta_{IA2}A_2)I,
\end{equation}

The lifetime of the infected cells is short. It depends on properties of the innate immune response. We set it to the time of virus reproduction, $\tau_I= 0.5$ days \cite{Timm2012}. $\beta_{IA1}$ and $\beta_{IA2}$  [ml/($\text{ng} \times \text{day}$)]  are the rates of destruction of infected cells following labeling by antibodies $A_1$ and $A_2$, respectively. Values were taken from the extended model. The antibodies are present at the time of infection due to vaccination. Note that in the model, $\beta_{IA1}$ and $\beta_{IA2}$ link tissue concentration of cells to body concentrations of antibodies.

\subsection[\appendixname~\thesubsection]{Virus}

Free-floating virus ($V$) can be taken up by susceptible cells that accordingly become infected, can be degraded or can be neutralized by antibodies A. As $V$ is shed into the lumen \cite{ChapuyRegaud2022}, its concentration is measured in copies per mL fluid. 

\begin{equation}
    \frac{\mathrm{d}V}{\mathrm{d}t} = p_V I -k_V \left(1-\frac{1}{I_{max}}I \right) SV - \frac{1}{\tau_V}V -(\beta_{VA1}A_1+\beta_{VA2}A_2)V,
\end{equation}

Here, $p_V$ is the virus production rate of infected cells. They can vary between 240 and 8000 copies/($\text{day} \times \text{cell}$) \cite{Timm2012}. We set: $p_V=2500$ copies/($\text{day} \times \text{cell}$). Assuming that one virus infects one cells i.e.  $k_V/k_{VI}=1$ copy/cell, one gets: $k_V=10^{-7}$  ml/($\text{day} \times \text{copy}$). We set the virus lifetime to: $\tau_V= 2.0$ days being at the lower bound of values found in respiratory secretions \cite{Guang2023}. $\beta_{VA1}$ and $\beta_{VA2}$ are virus neutralization rates by antibodies $A_1$ and $A_2$ respectively.

\subsection[\appendixname~\thesubsection]{Analytical solution for the critical titer $T_c$}

We first set $\beta_{IA1}A_1 + \beta_{IA2}A_2 = \beta_{IA}T$ and $\beta_{VA1}A_1 + \beta_{VA2}A_2 = \beta_{VA}T$ (see equations \eqref{eq:appbetaIA} and \eqref{eq:appbetaPfA}). For fast regeneration ($S=S_0$) and slowly evolving infection ($\mathrm{d}I/\mathrm{d}t = 0$), it follows:
\begin{equation}
    \begin{split}
        \frac{\mathrm{d}V}{\mathrm{d}t} &= p_VI - \left[ \frac{k_V}{k_{VI}} + \frac{\frac{1}{\tau_V}+\beta_{AV}T}{\left(1+\frac{1}{I_{max}}I\right) k_{VI}S_0} \right] \left( \frac{1}{\tau_I}+\beta_{AI}T \right) I \\
        &= (p_V-p_0)I.
    \end{split}
\end{equation}

The virus population vanishes for:  $\mathrm{d}V/\mathrm{d}t<0$, which is observed for: $p_V<p_0$. The virus production rate that can be tolerated without antibody for a starting infection ($I=0$) is:

\begin{equation}
    p_{VC} = \left[\frac{k_V}{k_{VI}}+\frac{\frac{1}{\tau_V}}{k_{VI}S_0} \right]\frac{1}{\tau_I}
\end{equation}

For the reference parameter set, one finds $p_{VC} =12$ copies/(day $\times$ cell), which is below reported rates.  For higher $p_V$, antibodies are required to control the infection. The critical titer is given by:

\begin{equation}
\begin{split}
    T_c &= -\frac{x}{2} + \sqrt{\frac{x^2}{2}-y} > 0, \quad \text{for: }p >p_{VC}, \\
    \text{with: } \quad x &= \frac{1}{\beta_{IA} \tau_I}+\frac{1}{\beta_{VA}}\left( \frac{1}{\tau_V} + k_V S_0 \right) > 0, \\
    y &= \frac{1}{\beta_{VA} \beta_{IA}} \left[ \left( \frac{1}{\tau_V} + k_V S_0 \right) \frac{1}{\tau_I} - pk_{VI}S_0 \right] <0.
\end{split}
\end{equation}

\begin{figure}[!htb]
    \centering
    \includegraphics[width=\linewidth]{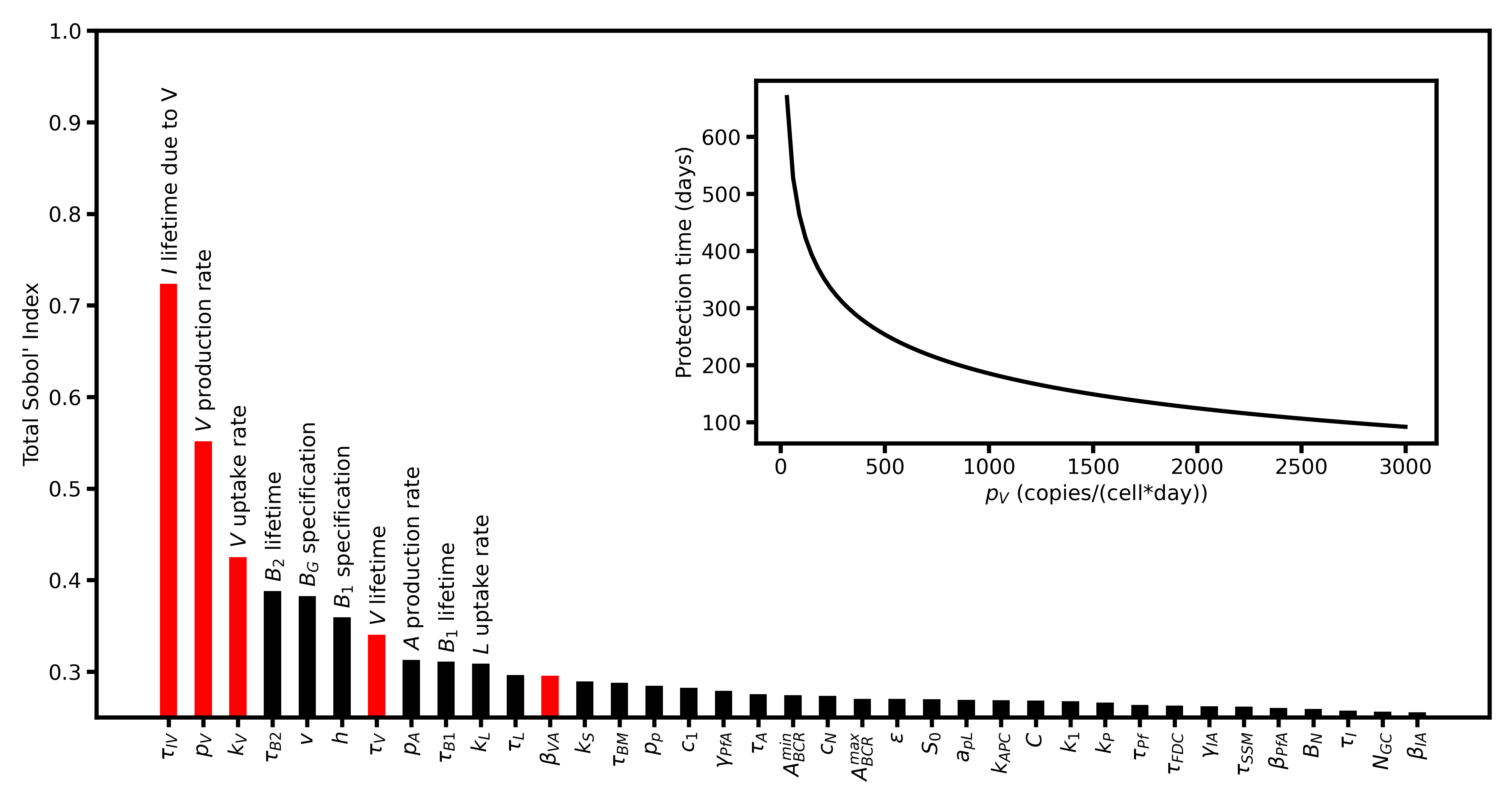}
    \caption{Sensitivity analysis I.  TSI regarding protection time. Black: Parameters of the extended model, Red: Parameters of the infection model. Insert: The protection time decreases with increasing virus replication rate $p_V$.}
    \label{fig:S2}
\end{figure}

\begin{figure}[!htb]
    \centering
    \includegraphics[width=\linewidth]{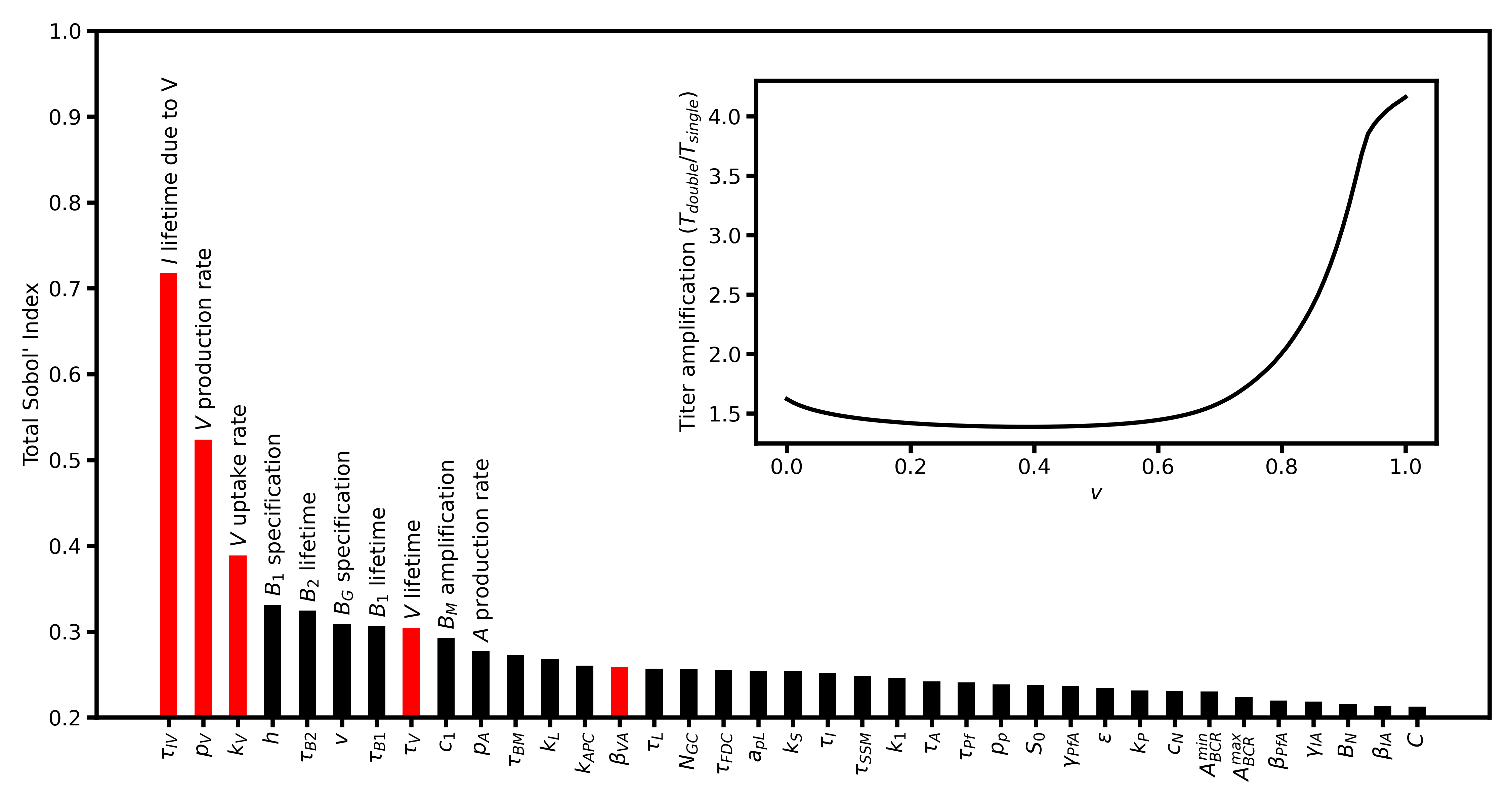}
    \caption{ Sensitivity analysis II. TSI for titer amplification by the second compared to the first dose. The parameters with the highest impact to the titer amplification are the lifetime of the virus infected cells $\tau_{IV}$, the virus production rate $p_V$ and the virus infection rate. Insert: For $v>0.5$, decreasing $B_2$ specification by increasing $v$ strengthen the titer amplification.}
    \label{fig:S3}
\end{figure}

$T_c$ depends on parameters of the virus infection model. We included these dependencies in our calculations of the parameter sensitivity for the protection time (Fig. \ref{fig:S2}). The titer amplification by the second dose (Fig. \ref{fig:S3}) does not depend on $T_c$.
\newpage

\newpage
\section[\appendixname~\thesection]{Reference parameter sets}

\begin{table}[!htb]
    \centering
    \begin{tabular}{|c|c|c|c|} \hline
        Parameter & Description & Value & Reference \\ \hline
        $k_L$ &  Uptake rate of $L$ & $2 \times 10^{-7}$ mL/day/cell & \cite{Kent2024} \\ \hline
        $\tau_L$ & Lifetime of $L$ & 7 days & \cite{Kent2024} \\ \hline
        $k_S$ & Infection rate & $2 \times 10^{-7}$ mL/day/cell & \cite{Kent2024} \\ \hline
        $\tau_I$ & Lifetime of $I$ & 4 days & \cite{Dalod2014} \\ \hline
        $k_1$ & Recruitment rate of $S$ & $7.1\times 10^{-2}$ /day & Set \\ \hline
        $S_0$ & $S$-pool density & $10^6$ cells/mL & \cite{Sender2023} \\ \hline
        $p_P$ & Protein production rate & 25 copies/day/cell & \cite{Sutton2023} \\ \hline
        $k_p$ & Internalization rate of $P_f$ & $10^{-7}$ mL/day/copy & See text \\ \hline
        $k_{APC}$ & Activation rate of APC &  $10^{-7}$ mL/day/copy & See text  \\ \hline
        $\tau_{Pf}, \tau_{PSSM}$ & Lifetime of $P_f$ and $P_{SSM}$ & 2 days & \cite{Cognetti2021} \\ \hline
        $\tau_{PFDC}$ & Lifetime of $P_{FDC}$ & 20 days & \cite{Turner2021} \\ \hline
    \end{tabular}
    \caption{Vaccination model}
    \label{tab:1a}
\end{table}

\begin{table}[!htb]
    \centering
    \begin{tabular}{|c|c|c|c|} \hline
        Parameter & Description & Value & Reference \\ \hline
        $B_N$ & $B_N$ density & $5 \times 10^4$ cells/mL & See text \\ \hline
        $N_{GC}$ & Number of productive GCs & 1 /mL & See text \\ \hline
        $c_N$ & Maximum $B_N$ recruitment rate & $10^{-3}$/day & \cite{Robert2024} \\ \hline
        $a_{pL}$ & GC apoptosis rate & 0.5/day & Set \\ \hline
        $\epsilon$ & Rate of $B_G$ leaving GC & 0.05/day & \cite{Robert2024} \\ \hline
        $A_{BCR}^{min}$ & Minimum $A_{BCR}$ & 1 & Set \\ \hline
        $A_{BCR}^{max}$ & Maximum $A_{BCR}$ & 25 & Set \\ \hline
        $v$ & Fraction of $B_G$ to $B_{MH}$ & 0.9 & Set \\ \hline
        $c_1$ & Maximum $B_M$ amplification rate & 0.6/day & \cite{Tangye2003} \\ \hline
        $h$ & Fraction of $B_M$ to $B_1$ & 0.6 & Set \\ \hline
        $\tau_{BM}$ & Lifetime of $B_M$ & 18 days & \cite{Macallan2005} \\ \hline
        $\tau_{B1}$ & Lifetime of $B_1$ & 5 days & \cite{Khodadadi2019} \\ \hline
        $\tau_{B2}$ & Lifetime of $B_2$ & 180 days & \cite{Manz1997} \\ \hline
        $p_A$ & Antibody production rate & 2 ng/day/cell & \cite{Bromage2009} \\ \hline
        $\tau_A$ & Antibody lifetime & 60 days & \cite{Barnes2021, Lau2022} \\ \hline
    \end{tabular}
    \caption{Response models}
    \label{tab:1b}
\end{table}

\begin{table}[!htb]
    \centering
    \begin{tabular}{|c|c|c|} \hline
        Parameter & Description & Value \\ \hline
        $C$ & Scaling rate & 0.1 cells/copy \\ \hline
        $\beta_{IA}$ & Cell destruction rate & $5 \times 10^{-5}$ mL/day/ng\\ \hline
        $\beta_{PfA}$ & Neutralization rate & $5 \times 10^{-5}$ mL/day/ng \\ \hline
        $\gamma_{IA}$ & Antibody consumption of $I$ & $2.5 \times \beta_{IA}$ ng/cell \\ \hline
        $\gamma_{PfA}$ & Antibody consumption of $P_f$ & $5 \times 10^{-4} \times \beta_{APf}$ ng/copy \\ \hline
    \end{tabular}
    \caption{Nonlinear terms and activation functions derived in Supplement \ref{supp:2}. }
    \label{tab:1c}
\end{table}

\begin{table}[!htb]
    \centering
    \begin{tabular}{|c|c|} \hline
        Population & Initial density \\ \hline
        $L$ & $2.5 \times 10^5$ copies/mL \cite{Kent2024} \\ \hline
        $S$ & $S_0$ \\ \hline
        $I, B_G, B_M, B_1, B_2$ & 0 \\ \hline
        $P_f, P_{FDC}, P_{SSM} $ & 0 \\ \hline
        $A_1,A_2$ & 0 \\ \hline
    \end{tabular}
    \caption{Initial densities of cells, antigens, and antibodies}
    \label{tab:init}
\end{table}

\begin{table}[!htb]
    \centering
    \begin{tabular}{|c|c|c|c|} \hline
        Parameter & Description & Value & Reference \\ \hline
        $k_{VI}$ & Cell infection rate & $10^{-7}$ mL/days/copy & Set \\ \hline
        $\tau_I$ & Lifetime of infected cells & 0.5 days & \cite{Timm2012} \\ \hline
        $S_0$ & $S$-pool density & $10^6$ cells/mL & Set \\ \hline
        $k_V$ & Virus internalization rate & $10^{-7}$ mL/days/cell & Set \\ \hline
        $p_V$ & Virus production rate & 2500 copies/day/cell & \cite{Timm2012} \\ \hline
        $\tau_V$ & Virus lifetime & 2 days & \cite{Guang2023} \\ \hline
    \end{tabular}
    \caption{Virus infection model: Parameters defining $T_c$.The rates of cell destruction, $\beta_{IA1}$ and $\beta_{IA2}$ and the rates of antigen neutralization $\beta_{VA1}$ and $\beta_{VA2}$ are taken from the response model ($\beta_{VA1} =\beta_{PfA1}$ and $\beta_{VA2}=\beta_{PfA2}$). }
    \label{tab:2}
\end{table}

\isPreprints{}{
} 

\reftitle{References}


\bibliography{main.bib}
%


\PublishersNote{}
\isPreprints{}{
} 
\end{document}